\documentclass[twocolumn,prl,superscriptaddress,floatfix]{revtex4-1}

\usepackage{verbatim}
\usepackage{textcomp}
\usepackage[official]{eurosym}
\usepackage{amsmath}
\usepackage{amssymb}
\usepackage{bm}

\usepackage[dvipsnames]{xcolor}
\usepackage{tikz}
\usetikzlibrary{arrows}
\AtBeginDocument{\usepackage{booktabs}}
\usepackage{hyperref}
\usepackage{bookmark}
\usepackage{braket}
\usepackage{siunitx}
\usepackage[version=4]{mhchem}
\usepackage{csquotes}

\newcommand{\rv}{\bm{{\rm r}}}

\newcommand{\Bv}{\bm{{\rm B}}}

\newcommand{\vv}{\bm{{\rm v}}}

\newcommand{\nv}{\bm{{\rm n}}}

\newcommand{\kv}{\bm{{\rm k}}}

\newcommand{\Lv}{\bm{{\rm L}}}
\newcommand{\jv}{\bm{{\rm j}}}

\newcommand{\qv}{{\bm{{\rm q}}}}

\newcommand{\qb}{{\bar{q}}}
\newcommand{\dd}{\text{d}}
\newcommand{\ii}{\text{i}}
\newcommand{\ZZ}{\mathbb{Z}}
\newcommand{\RR}{\mathbb{R}}
\def\ee{\text{e}}
\def\ii{\text{i}}
\def\nuo{\nu^{\text{o}}}

\def\sign{\textrm{sign}}

\begin{document}

\title{Topological waves in fluids with odd viscosity}

\author{Anton Souslov}
\email{a.souslov@bath.ac.uk}
\affiliation{The James Franck Institute,
The University of Chicago, Chicago, IL 60637, USA}
\affiliation{Department of Physics, University of Bath, Bath BA2 7AY, United Kingdom}

\author{Kinjal Dasbiswas}
\affiliation{The James Franck Institute,
The University of Chicago, Chicago, IL 60637, USA}
\affiliation{Department of Physics, University of California, Merced, Merced, CA 95343, USA}

\author{Michel Fruchart}
\affiliation{The James Franck Institute,
The University of Chicago, Chicago, IL 60637, USA}

\author{Suriyanarayanan Vaikuntanathan}
\affiliation{The James Franck Institute,
The University of Chicago, Chicago, IL 60637, USA}
\affiliation{Department of Chemistry,
The University of Chicago, Chicago, IL 60637, USA}

\author{Vincenzo Vitelli}
\email{vitelli@uchicago.edu}
\affiliation{The James Franck Institute,
The University of Chicago, Chicago, IL 60637, USA}
\affiliation{Department of Physics, 
The University of Chicago, Chicago, IL 60637, USA}

\begin{abstract}
\noindent
Fluids in which both time-reversal and parity are broken can display a dissipationless viscosity that is odd under each of these symmetries. Here, we show how this odd viscosity has a dramatic effect on topological sound waves in fluids, including the number and spatial profile of topological edge modes. Odd viscosity provides a short-distance cutoff that allows us to define a bulk topological invariant on a compact momentum space. As the sign of odd viscosity changes, a topological phase transition occurs without closing the bulk gap. Instead, at the transition point, the topological invariant becomes ill-defined because momentum space cannot be compactified. This mechanism is unique to continuum models and can describe fluids ranging from electronic to chiral active systems.
\end{abstract}

\date{\today}
\maketitle

In ordinary fluids, acoustic waves with sufficiently large wavelength have arbitrarily low frequency due to Galilean invariance~\footnote{The equations of motion of the fluid are invariant under the Galilean transformations $\vv(x,t) \rightarrow \vv(x,t) + \mathbf{v}_0$, where $\vv(x,t)$ is the velocity field of the fluid particles and $\mathbf{v}_0$ an arbitrary uniform velocity field corresponding to an inertial frame. Hence, density waves with arbitrarily large wavelength (approaching a uniform flow) have vanishingly small frequency.}.
When either a global rotation or an external magnetic field is present, Galilean invariance is explicitly broken by either Coriolis or Lorentz forces within the fluid, respectively. Hence, the spectrum of acoustic waves becomes gapped in the bulk. Yet, a peculiar phenomenon can occur at edges or interfaces: chiral edge modes propagate robustly irrespective of interface geometry. This phenomenon analogous to edge states in the quantum Hall effect~\cite{Klitzing1980,Thouless1982,Hasan2010} was unveiled in the context of equatorial waves~\cite{Delplace2017} and explored in out-of-equilibrium and active fluids~\cite{Shankar2017,Dasbiswas2017}. Similar phenomena occur in lattices of circulators~\cite{Fleury2014,Khanikaev2015}, polar active fluids under confinement~\cite{Souslov2017} and coupled mechanical oscillators~\cite{Huber2016,Susstrunk2016,Susstrunk2015}, including gyroscopes~\cite{Nash2015,Wang2015b} and oscillators subject to Coriolis forces~\cite{Wang2015a,Kariyado2015}.

In addition to Coriolis or Lorentz body forces, fluids in which time-reversal and parity are broken generically exhibit a dissipationless viscosity that is odd under each of these symmetries~\cite{DeGrootMazur,Avron1998}. The viscosity tensor $\eta_{ijkl}$ relates the strain rate $v_{kl} \equiv \partial_k v_l$ to the viscous part of the stress tensor $\sigma_{ij} = \eta_{ijkl} v_{kl}$. Odd viscosity refers to the antisymmetric part of the viscosity tensor $\eta^{\text{o}}_{ijkl} = - \eta^{\text{o}}_{klij}$~\cite{DeGrootMazur,Avron1998}. In an isotropic two-dimensional fluid, odd viscosity is specified by a single pseudoscalar $\eta^{\text{o}}$, see Supplementary Information (SI) for details~\cite{Avron1998}. Odd viscosity changes sign under either time-reversal or parity, and hence must vanish when at least one of these symmetries is present. Conversely, odd viscosity is generically non-vanishing as soon as both time-reversal and parity are broken \cite{Jensen2012,Kaminski2014,Haehl2015}. For instance, microscopic Coriolis or Lorentz forces are sufficient to induce a non-zero odd viscosity \cite{Nakagawa1956,ChapmanCowling}, in addition to the corresponding body forces. Odd viscosity has been studied theoretically in various systems (see SI for a partial review) including polyatomic gases~\cite{Knaap1967}, magnetized plasmas~\cite{Landau10,ChapmanCowling}, fluids of vortices~\cite{Wiegmann2014,Bogatskiy2018,Abanov2018a,Abanov2018b}, chiral active fluids~\cite{Banerjee2017}, quantum Hall states and chiral superfluids/superconductors~\cite{Volovik1984,Avron1995,Tokatly2007,Tokatly2009,Read2009,Read2011,Barkeshli2012,Shitade2014,Vollhardt,Fujii2018,Offertaler2019}. Its presence has been experimentally reported in polyatomic gases~\cite{Korving1966,Korving1967,Hulsman1970} (where both positive and negative odd viscosities were observed under the same magnetic field, for different molecules), electron fluids subject to a magnetic field~\cite{Berdyugin2018}, and spinning colloids~\cite{Soni2018}.

\begin{figure}
	\includegraphics[angle=0]{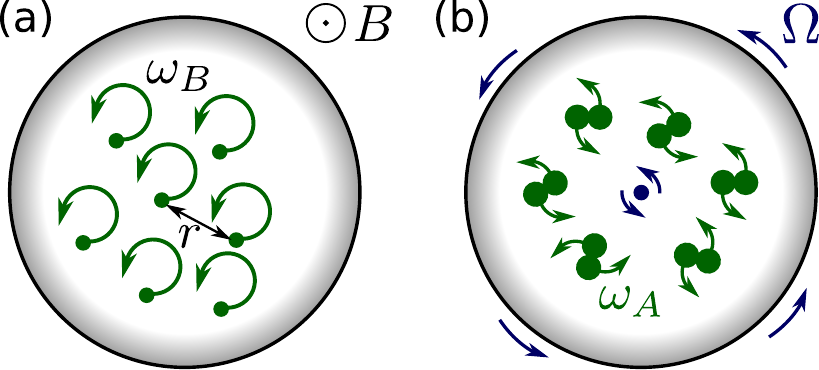}
	\caption{
	Physical realizations of the minimal model for topological fluids with odd viscosity. 
	(a) Two-dimensional plasma under magnetic field $B$, with cyclotron frequency $\omega_B = q B/M$.
	(b) Chiral active fluid with intrinsic rotation angular frequency $\omega_A$,
	subject to a global rotation with angular frequency $\omega_B = -2 \Omega$.
      } 
	\label{Fig1}
\end{figure}

Here, we show that the presence of odd viscosity fundamentally affects the topological properties of linear waves in the fluid. In particular, the net number of chiral edge states depends on the signs of both odd viscosity and the external magnetic field (or rotation) on each side of an interface. We define a bulk topological invariant that accounts for this striking behavior. 
In a fluid, momentum space is not compact (linear momentum can be arbitrarily large). Hence, the definition of bulk topological invariants requires a constraint at short wavelengths~\cite{So1985,Volovik1988,Coste1989}.
We show that a non-vanishing odd viscosity provides such a short-distance cutoff, associated with microscopic angular momenta (see Fig.~\ref{Fig1}). Upon changing the sign of odd viscosity, a topological phase transition occurs without gap closing because at the transition, the small-wavelength constraint changes, so the topological invariant becomes ill-defined. 
When odd viscosity goes to zero, half of the edge states are no longer hydrodynamic because their penetration depths vanish while the other half retain a finite penetration depth set by the gap size.

\emph{Model}.--- 
Consider the {odd Navier-Stokes equations describing a} compressible time-reversal and parity violating fluid:
\begin{align}
\label{eq:cont}
\partial_t \rho(\rv,t) & = - \rho_0 \nabla \cdot \vv(\rv,t) \\
\partial_t \vv & = - c^2 \nabla \rho/\rho_0 + \omega_B \vv^* + \nuo \nabla^2 \vv^*
\label{eq:mom}
\end{align}
where $\rv \equiv (x,y)$ is the position, $\rho$ is the fluid density whose average is $\rho_0$~\footnote{The odd Navier-Stokes equations are linearized around the state $(\rho, \vv) = (\rho_0, 0)$. In a rotating fluid, $\vv$ is the velocity in the rotating frame and this state corresponds to rigid-body rotation.}, $\vv \equiv (v_x, v_y)$ is the velocity and $\vv^*\equiv (v_y, - v_x)$ is the velocity rotated by \ang{90}. The chiral body force $\omega_B \vv^*$ can arise, e.g., from (i) Lorentz forces for which $\omega_B = q B/M$ where $B \hat{z}$ is the magnetic field, $q$ is particle charge, and $M$ is particle mass or (ii) Coriolis forces for which $\omega_B = - 2 \Omega$ where $\Omega \hat{z}$ is the rotation field. Besides the body force $\omega_B \vv^*$, the Lorentz or Coriolis forces experienced by the fluid particles also give rise to an odd viscosity term $\nuo \nabla^2 \vv^*$, see SI and Ref.~\cite{Nakagawa1956,ChapmanCowling} for kinetic theory derivations and the dependence of odd viscosity on fluid parameters, including temperature. Other microscopic mechanisms violating both time-reversal and parity also contribute to the odd viscosity. This is for instance the case of active torques (see SI and Ref.~\cite{Banerjee2017}).

Equations~(\ref{eq:cont}-\ref{eq:mom}) are our starting point. Equivalent equations, but with zero odd viscosity and with $\rho$ replaced by the height of a surface wave, are studied in the context of geophysics~\cite{Thomson1880,Rosenthal1965,Matsuno1966,Yanai1966,Maruyama1967,Gill1982}. 
The topological properties of such waves were identified for fluids on a sphere~\cite{Delplace2017}, see also Ref.~\cite{Shankar2017}.
In the next section, we show that a non-zero odd viscosity allows the topological characterization of density waves for fluids within a plane by acting as a short distance cut-off, see also Ref.~\footnote{While this paper was under review, we became aware of the preprint Ref.~\cite{Tauber2018}, in which the effect of odd viscosity on geophysical waves is discussed.}. By contrast, an ordinary viscosity term $\nu \nabla^2 \vv$ by itself does not lead to a regularization of the continuum theory; this term can be neglected in the limit $\nuo/\nu \gg 1$ (see SI for a discussion).

\begin{figure}
	\includegraphics[angle=0]{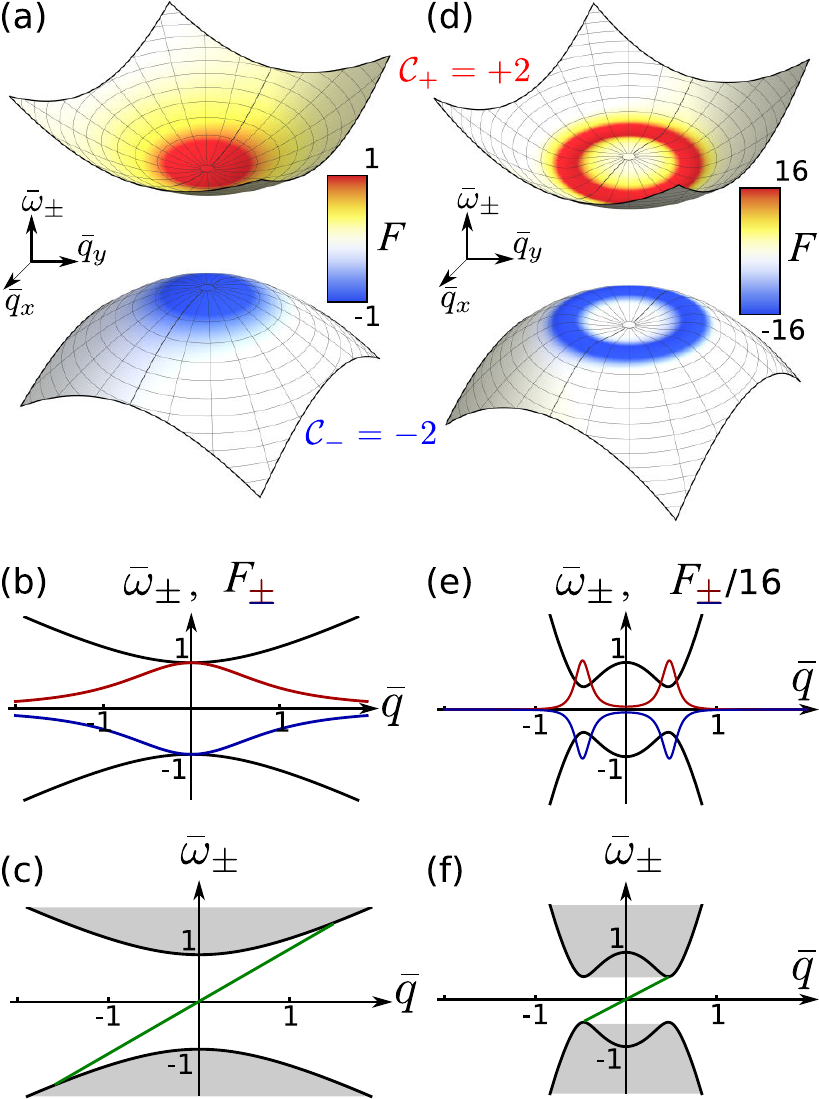}
	\caption{
	Effect of large odd viscosity on the topological band structure.
	(a) Frequency $\bar{\omega}_\pm \equiv \omega_\pm(\qv)/\omega_B$ and Berry curvature $F_\pm$ 
	for $m = 0.1$.
	(c) Schematic representation of band structure
	for system with edge. Gray regions of bulk states are connected by lines of edge states.
	(d--f) Same as (a--c), but for $m = 4$.}
	\label{Fig2}
\end{figure}

\emph{Bulk dispersion and topology}.--- 
In the fluid bulk, Eqs.~(\ref{eq:cont}-\ref{eq:mom}) can be replaced by their momentum-space version
$	\partial_t [\rho, \vv] = \ii \mathcal{L}(\qv) [ \rho , \vv ]	$
where the operator $\mathcal{L}(\qv) \equiv q_x \Lambda_x + q_y \Lambda_y + (\omega_B - \nuo q^2) \Lambda_z$ is expressed in terms of the $3 \times 3$ matrices $\Lambda_i$ ($i = x,y,z$, see SI for definitions). Here, $\rho(\qv,t), \vv(\qv,t)$ are the Fourier transforms of $\rho(\rv,t), \vv(\rv,t)$, and the wavevector $\qv \equiv (q_x,q_y)$ takes values in the entire plane. The dispersion relations $\omega_s(\qv)$ for the frequency of bulk modes are the eigenvalues of $\mathcal{L}(\qv)$ and consist of three branches.
One branch has a flat dispersion $\omega_0(\qv) = 0$ with an eigenmode combining vorticity and density (see SI).
The acoustic spectrum is described by the other two branches, with dispersion relations
\begin{equation}
\omega_\pm(\qv)/\omega_B = \pm \sqrt{(1 - m \qb^2)^2 + \qb^2} , \label{eq:disp-b}
\end{equation}
where $\qb = |\qv| c/\omega_B$.
The qualitative features of these dispersion relations near $\qb = 0$ depend on the frequency $\omega_B$ and the dimensionless velocity ratio $m \equiv  \omega_B \nuo/c^2$, which is analogous to the square of a Mach number (see SI).
As $\omega_B$ (and not $\nuo$) controls the magnitude of the gap at $\qb = 0$,
odd viscosity alone cannot open a gap in the spectrum of acoustic waves.
However, $m$ plays an important role in the shape of the dispersion relation. 
For $m < 1/2$, the band structure looks similar to the case $m = 0$, see Fig.~\ref{Fig2}a--c.
For $m > 1/2$, the band structure resembles a Mexican-hat potential. While the separation between the bands is unchanged at $\bar{q}=0$, the gap is now located along a circle with radius $\bar{q} = \mathrm{const} \ne 0$, and the gap size decreases scaling as $\omega_B m^{-{1/2}}$ at large $m$.
In this regime, the group velocity $\partial \omega_+ / \partial q$ of sound waves in the fluid is negative for $0 < \qb < \sqrt{(2 m - 1)/ (2 m^2)}$, a feature shared with left-handed metamaterials, which have a negative index of refraction.

The analogy between acoustic waves on top of a constant background vorticity and the quantum-mechanical wavefunction of electrons in a constant magnetic field suggests that Eqs.~(\ref{eq:cont}-\ref{eq:mom}) can lead to topological phenomena akin to the quantum Hall effect. 
The geometric phases in the wave propagation are captured by the Berry curvature $F_{\pm}(\qv) = \nabla_\qv \times [(u^\pm_\qv)^\dagger \cdot \nabla_\qv u^\pm_\qv]$ of the eigenmodes $u^\pm_\qv$ associated with the $\pm$ bands at $\qv$ in Eq.~(\ref{eq:disp-b}), that reads
\begin{equation}
F_{\pm}(\bar{q}) = \mp \frac{1 + m \bar{q}^2}{\big[\bar{q}^2 + (1 - m \bar{q}^2)^2\big]^{3/2}}.
\label{eq:b1}
\end{equation}

In the usual case, the integral of Berry curvature over momentum space is equal to a topological invariant.
However, standard topological materials have a lattice structure, for which the wavevector $\qv$ lives in a compact Brillouin zone, equivalent to a torus.
In contrast, fluid models such as the one described by Eqs.~(\ref{eq:cont}-\ref{eq:mom}) 
do not include a short-distance cutoff, and the wavevector spans the entire two-dimensional $(q_x,q_y)$ plane. 
As a consequence, the definition of topological invariants for fluid models requires the introduction of a constraint at small length scales~\cite{So1985,Volovik1988,Coste1989,Volovik2009,Volovik2009b,Silaev2014,Li2010,Silveirinha2015,Silveirinha2016,Bal2018}, resulting in a nonzero $m$ in Eq.~\eqref{eq:b1}. Formally, this addition can be seen as an ultraviolet regularization of the continuum model. Here, a mesoscopic length scale naturally arises from odd viscosity whose presence leads to a well-defined limit for $\mathcal{L}(\qv)$ as $|\qv| \to \infty$, independent of the direction of $\qv$. As a result, integer-valued topological invariants can be associated to each band of the wave spectrum as the first Chern numbers of a modified version of the operator $\mathcal{L}$ defined over the compactified momentum space, i.e. a sphere (see SI and Refs.~\cite{So1985,Volovik1988,Coste1989,Volovik2009,Volovik2009b,Silaev2014,Li2010,Silveirinha2015,Silveirinha2016,Bal2018} in which a different short-distance cutoff is considered in other physical contexts).

When both $\omega_B$ and $\nuo$ are nonzero (and only in this case), the first Chern number~$\mathcal{C}_{-}$ of the band with dispersion~$\omega_{-}$ is given by
\begin{equation}
	\label{chern_number}
	\mathcal{C}_{-} = \sign(\nuo) + \sign(\omega_B),
\end{equation}
whereas the other acoustic band has the opposite first Chern number $\mathcal{C}_{+} = - \mathcal{C}_{-}$, 
and the flat band $\omega = 0$ has a vanishing first Chern number.
When odd viscosity vanishes, $\mathcal{L}(\qv)$ does not have a unique limit as $|\qv| \to \infty$. Hence, the compactification is no longer possible, and the Chern numbers become ill-defined.
Remarkably, this results in a topological phase transition without gap closing \cite{Volovik1988,Volovik2009b,Silaev2014}.
This phase transition is due to an ultraviolet divergence of the hydrodynamic field theory. In other words, the hydrodynamic description of the system breaks down as the small lengthscales associated with odd viscosity vanish. 

The distribution of Berry curvature is also qualitatively modified by odd viscosity (see Fig.~\ref{Fig2}).
When $0 < m < 3/8$, the Berry curvature concentrates at $\bar{q}=0$. At higher values $m > 3/8$, the Berry curvature concentrates on a ring with finite radius, scaling as $\bar{q} \sim m^{-{1/2}}$ for large $m$. For negative $m$, a peak at $\bar{q}=0$ coexists with an extremum along a ring, with opposite contribution.

\begin{figure}
	\includegraphics[angle=0]{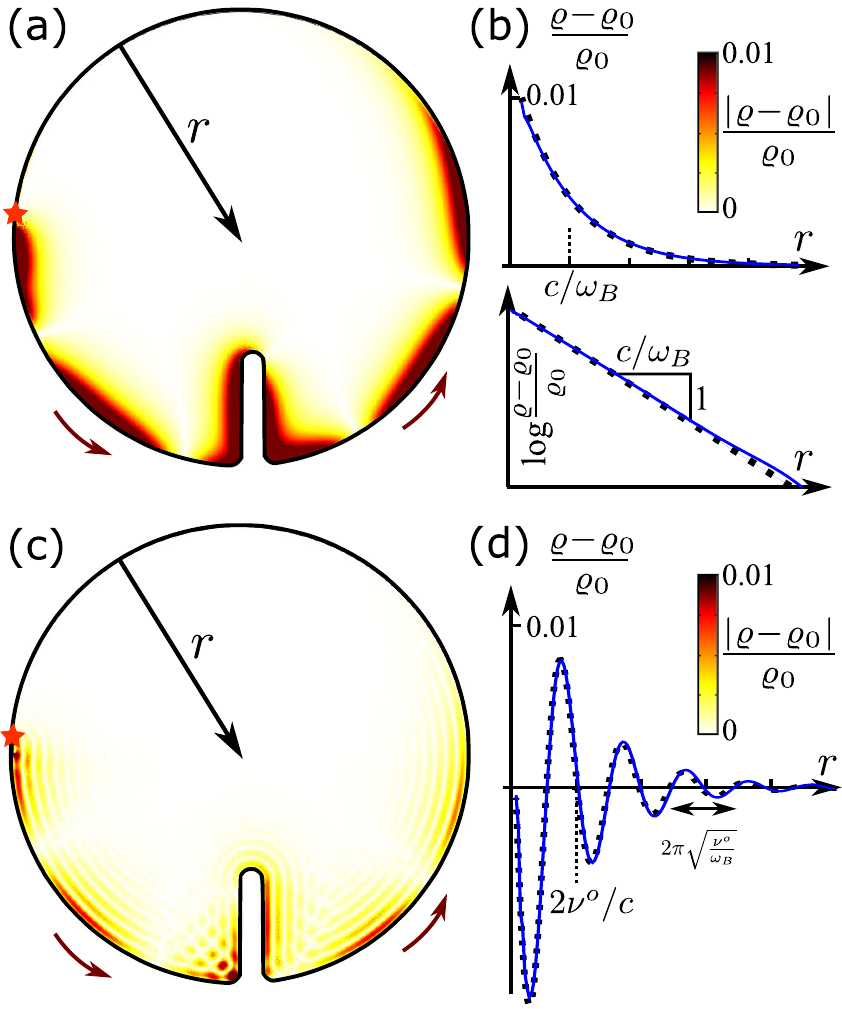}
	\caption{
 Simulations showing topological edge states. (See Supplementary Movie)
    (a) Edge state with $m = 0.0625$. 
    Color shows density deviations, $|\rho - \rho_0|/\rho_0$.
    The edge state is excited using small-frequency source (star, left).
    (b, top) Radial profile of the chiral edge state in (a) comparing simulations and analytics.
    (b, bottom) Same plot on log-linear scale. 
    (c) Snapshot as in (a), but with $m = 9.0$.
    (d) Radial profile from (c) exhibits oscillations.
    }
	\label{Fig3}
\end{figure}

\emph{Bulk-boundary correspondence.}--- 
Topological invariants characterize infinite systems without boundaries, 
but their values are usually related to observable phenomena at interfaces.
According to bulk-boundary correspondence, the net number of chiral edge states (with frequencies in the bulk band gap) expected at an interface between two systems L and R with invariants $\mathcal{C}_{-}^{\text{L/R}}$, respectively, is $N=\mathcal{C}_{-}^{\text{L}} - \mathcal{C}_{-}^{\text{R}}$.
Note that the general validity of bulk-boundary correspondence has not been established in continuum fluid models.
{We assume that the} case of a container wall can be considered by setting $\mathcal{C}_{-}^{\text{R}} = 0$ for the region where waves cannot propagate \cite{Schnyder2009,Hasan2010}.
Provided that both $\omega_B$ and $\nuo$ are nonzero, Eq.~\eqref{chern_number} applied to the region where waves propagate implies that a chiral fluid has a total of two protected edge modes traveling in the same direction at an edge if $\omega_B \nuo > 0$
(corresponding to $|\mathcal{C}^{\text{L}}_{-}| = 2$), or a net total of zero chiral edge modes if $\omega_B \nuo < 0$ (corresponding to $\mathcal{C}^{\text{L}}_{-} = 0$).
Notably, a topological phase transition occurs between these two regimes without closing the bulk band gap.
Here, the second case corresponds to two counter-propagating edge states which are not topologically protected (see SI and Supplementary Movie). 
We demonstrate these phenomena within finite-element simulations
of Eqs.~(\ref{eq:cont}--\ref{eq:mom}) in a modified disk geometry using COMSOL Multiphysics
(see Fig.~\ref{Fig3}, SI, and Supplementary Movies).
The density wave at the edge is excited at a frequency in the gap (c.f., Fig.~\ref{Fig2}).
For a range of model parameters with $\omega_B \nuo > 0$, 
the edge waves propagate unidirectionally around the edge of the disk and do not scatter off sharp corners and prominent defects.
Similarly, an interface between fluids with opposite $\omega_B$ with $\omega_B \nuo > 0$ on both sides should exhibit four co-propagating edge states. 
This is in sharp contrast to the case of strictly vanishing odd viscosity \cite{Delplace2017,Shankar2017,Dasbiswas2017}, where only two edge modes are present at an interface.

Although the existence of chiral edge states relies
only on the nonzero topological invariant associated with the bulk bands, 
their penetration depth is determined by the various parameters in Eqs.~(\ref{eq:cont}-\ref{eq:mom}).
The penetration depth depends on the separation between the two topological bands, which can scale with odd viscosity.
To estimate this penetration depth $\kappa^{-1}$, we consider a simplified geometry with a straight fluid interface perpendicular to the $y$-axis 
with a fluid described by Eqs.~(\ref{eq:cont}-\ref{eq:mom}) filling the region $y < 0$, whereas the region $y > 0$ is empty.
Along this edge, solutions for density waves in the fluid have the form $\ee^{\ii (\omega t - \qv\cdot\rv) + \kappa y}$ (for $y < 0$), 
which decays to zero as $y \rightarrow - \infty$ for real $\omega$, $q_x$, $q_y$ and positive $\kappa$. 
We assume that the dispersion
of the edge states goes through the point $\omega(q_x = 0) = 0$ (see SI for the general case).
From Eq.~(\ref{eq:disp-b}) where $\bar{\kappa} \equiv \kappa c/\omega_B$, we find
\begin{equation}
 \left[ \big[1 - m (\bar{q}_y + \ii  \bar{\kappa})^2\big]^2 
 +  (\bar{q}_y + \ii  \bar{\kappa})^2\right]^{1/2} = 0. 
\label{eq:disp-edge}
\end{equation}
For small odd viscosity with $0 < m < 1/4$, we find solutions with $\bar{q}_y = 0$
and $\kappa_{\pm} = (c \pm \sqrt{c^2 - 4 \nuo \omega_B})/2 \nuo$.
This solution includes the case $\kappa_{-} \to \omega_B/c$ in the limit $\nuo \rightarrow 0$~\cite{Delplace2017}.
In this limit, $\kappa_+ \sim c/\nuo \to \infty$ which implies this mode has vanishing penetration depth and therefore is no longer hydrodynamic.
By contrast, no solution satisfying $q_y = 0$ exists at large odd viscosity when $m > 1/4$. 
Instead, the edge wave has a profile whose amplitude both decays and oscillates away from the edge.
When $m \gg 1$ this solution has the form $q_y = \pm \sqrt{\omega_B/\nuo}$
and $\kappa = c/ (2 |\nuo|) \sim m^{-1} \omega_B/c \ll \omega_B/c$.
In Fig.~\ref{Fig3}, we compare these results with numerical simulations in which no-tangential-stress, no-penetration boundary conditions have been chosen (see SI for details, where we also observe that no-slip boundary conditions do not lead to qualitative changes).
We find good agreement between our theoretical predictions and numerical simulations for
both the penetration depth (for $\nuo$ both small, Fig.~\ref{Fig3}b, and large, Fig.~\ref{Fig3}d)
and the oscillation wavelength (for large $\nuo$, Fig.~\ref{Fig3}d).

\emph{Discussion}.---
When can odd viscosity be neglected? Comparing the magnitudes of terms on the right hand side of Eq.~(\ref{eq:mom}), we find two length scales $\ell_1 \equiv \nuo/c$ and $\ell_2 \equiv  \sqrt{\nuo/\omega_B}$ from the ratio of the compressibility and Lorentz or Coriolis terms to odd viscosity (see SI). At scales significantly larger than $\ell_{1,2}$, odd viscosity is a small effect. When $\nuo \to 0$, $\ell_{1,2}$ both vanish and the effects of odd viscosity are no longer captured by the hydrodynamic description. In this case, the lack of a cutoff at short wavelength in the band structure allows for a topological phase transition without a corresponding closing of the band gap. In a topological system with boundaries, the penetration depth of one of the edge states scales as $\ell_1$ in the limit $\nuo \to 0$, whereas the penetration depth of the other edge state converges to a finite value. In this limit, the effects of odd viscosity are confined in a boundary layer with small thickness of order $\ell_1$, in which the hydrodynamic description does not apply. In particular, we find that in the limit of vanishing odd viscosity, a single edge state with finite penetration depth remains, with a chirality controlled by the sign of $\omega_B$. The other edge state with vanishingly small penetration depth is either co-propagating or counter-propagating, depending on the relative sign of $\omega_B$ and $\nuo$, but likely becomes unobservable in the limit of zero odd viscosity, in agreement with the results of Refs.~\cite{Delplace2017,Shankar2017,Dasbiswas2017}.

When the lengthscales associated with odd viscosity are sufficiently large, both edge states should be observable, and different signs of odd viscosity relative to $\omega_B$ lead to physically distinct situations. 
Positive and negative $\omega_B \nuo$ are possible even when the body force and odd viscosity both arise from the same origin.
For instance, polyatomic gases under magnetic field can have an odd viscosity of either sign in the same magnetic field, depending on the constituent molecules~\cite{Korving1966,Korving1967,Hulsman1970}.  
Besides, active systems may allow one to control both quantities independently due to an additional internal source of time-reversal and parity violation. For example, chiral active fluids consist of microscopic components of size $a$ subject to internal torques and dissipation~\cite{Dahler1961, Condiff1964, Tsai2005,Bonthuis2009,Furthauer2012, Sumino2012, Tabe2003, Oswald2015, Drescher2009, Petroff2015, Yan2015, Riedel2005, Denk2016, Snezhko2016, Maggi2015, Lemaire2008, Lenz2003, Uchida2010, Yeo2015, Spellings2015, Nguyen2014, vanZuiden2016, Han2017},
resulting in a steady-state rotation of each microscopic component with frequency $\omega_A$ and 
an odd viscosity $\nuo \propto \omega_A a^2$~\cite{Banerjee2017}, where $\omega_A$ and $\omega_B$ can have opposite signs.

\begin{acknowledgments}
	\emph{Acknowledgments}.--- We thank Alexander Abanov, Guillaume Bal, Eric Eliel, Andrey Gromov, William Irvine, Tom Lubensky, 
Noah Mitchell, Paul Wiegmann, and Tom Witten for fruitful discussions.
This work was primarily supported by the University of Chicago Materials Research Science and Engineering Center, which is funded by the National Science Foundation under award number DMR-1420709.
\end{acknowledgments}

\clearpage
\begin{center}
\noindent \textbf{\uppercase{Supplementary Information}}
\end{center}

\setcounter{equation}{0}
\setcounter{figure}{0}

\renewcommand{\theequation}{S\arabic{equation}}
\renewcommand{\figurename}{{\bf Fig. }}
\renewcommand{\thefigure}{{\bf S\arabic{figure}}}

\makeatletter
\c@secnumdepth=4
\makeatother

\section{Derivation of the equations of motion}
In this section, we derive Eqs.~(1--2) of the main text to describe a two-dimensional magnetized plasma with a Lorentz force, a standard compressible fluid under rotation, or a chiral active fluid with a Coriolis force.

\subsection{Linearization of the pressure term}

We consider perturbations of the fluid around a steady state in which the density, velocity, and pressure fields are given by $(\rho_0, \vv_0, p_0)$, respectively.
For waves with small amplitudes, the equation of state $p(\rho)$ can be expanded as
\begin{equation}
	p(\rho) = p_0 + \left.\frac{\partial p}{\partial \rho}\right|_{\rho_0} (\rho - \rho_0) + \mathcal{O}(\rho - \rho_0)^2.
\end{equation}
where the response $c^2 = (\partial p/\partial \rho)_{\rho_0}$ is by definition the square of the local speed of sound $c$ (it is assumed that the derivative is taken at constant entropy). Hence, we obtain at first order
\begin{equation}
	\label{pressure_density}
	\nabla p = c^2 \nabla \rho.
\end{equation}

\subsection{Magnetized one-component thermal plasmas with screened interactions}

We start with the Navier-Stokes equations in two dimensions
\begin{align}
\partial_t \rho(\rv,t) + \nabla \cdot [\rho \vv(\rv,t)] = 0 \label{eom_main_hydro_rho} \\
\rho \partial_t \vv + \rho (\vv \cdot \nabla) \vv = - \nabla p + \jv \times \Bv + \eta^{\text{o}} \nabla^2 \vv^*
\label{eom_main_hydro_v}
\end{align}
To account for the effect of the external magnetic field $\Bv$, a Lorentz body force term $\jv \times \Bv$ and the anomalous odd viscosity term $\eta^{\text{o}} \nabla^2 \vv^*$ are considered (see Refs.~\cite{Landau10, Korving1966, Hulsman1970} for the equations of fluid dynamics and transport coefficients in magnetized plasmas), where $\jv = q/m \rho \vv$ is the current density and $\Bv$ the magnetic field. Here, we consider a magnetic field orthogonal to the plane, so the Lorentz term becomes $\rho \omega_B \vv \times \hat{z} = \rho \omega_B \vv^*$.
We consider the regime in which viscous dissipation can be ignored, and do not include an ordinary viscosity term $\eta \nabla^2 \vv$.
Finally, we linearize the equations around the steady state $(\rho,\vv) = (\rho_0,0)$ and replace the pressure term through Eq.~\eqref{pressure_density} to obtain Eqs.~(1--2).

\subsection{Rotating fluid}

We now consider a fluid undergoing solid-body rotation with angular frequency $\omega_R$ in the steady state. 
In the rotating frame, a microscopic Coriolis force acts on the constituents of the fluids, giving rise to an odd viscosity term~\cite{Nakagawa1956} and a Coriolis body force. 
The Coriolis body force can be obtained by linearizing the convective derivative around the rotating flow field.
The change of reference frame to the co-rotating frame leads to the transformation $\partial_t \vv \rightarrow \partial_t \vv - \omega_R \vv^*$, while the linearization of the convective term leads to $(\vv \cdot \nabla) \vv \rightarrow - \omega_R \vv^*$.
Hence, we end up with the equations
\begin{align}
\partial_t \rho(\rv,t) + \nabla \cdot [\rho \vv(\rv,t)] = 0 \\
\rho \partial_t \vv + \rho (\vv \cdot \nabla) \vv  = - \nabla p + 2 \rho \omega_R \vv^* + \eta^{\text{o}} \nabla^2 \vv^*.
\end{align} 
In this equation, $\vv$ is the velocity field in the rotating frame. By linearizing around $(\rho,\vv) = (\rho_0,0)$, i.e. around the solid-body rotation, and replacing the pressure term through Eq.~\eqref{pressure_density}, we again obtain Eqs.~(1--2).

\subsection{Chiral active fluids}
Now we consider the derivation of a similar set of equations describing
the hydrodynamics of a fluid which consists of particles spinning around their centers---a chiral fluid with active torques.
We show that rigid-body rotation can arise in such a fluid
as a result of the combination of the active torques with a specific choice of boundary conditions.
When the velocity field describes a rigid-body flow, 
the equations describing density waves in these chiral active fluids are Eqs.~(1--2) of the main text.
The full nonlinear equations of motion for such a fluid includes
three dynamical fields: the density $\rho(\rv,t)$, velocity $\vv(\rv,t)$, and local intrinsic
rotation rate $\Omega(\rv,t)$~\cite{Tsai2005,vanZuiden2016,Banerjee2017}:
\begin{gather}
	\partial_t\rho(\rv,t) +\partial_i( \rho(\rv,t) v_i(\rv,t) ) = 0
 \label{eq-rho-nl} \\
	\partial_t (I \Omega) +\partial_i(I \Omega v_i) = \tau+D^\Omega \partial_i^2\Omega
	- \Gamma^\Omega \Omega
	- \epsilon_{ij}\sigma_{ij}
 \label{eq-om-nl} \\
	\partial_t (\rho v_i) + \partial_j(\rho v_i v_j ) = \partial_{j}\sigma_{ij}-\Gamma^v v_i
 \label{eq-v-nl} 
\end{gather}
where $I \equiv \iota \rho$ is the moment of inertia density ($\iota \sim a^2$, where $a$ is the linear size of the fluid's constituents particles),
$\tau$ is the density of external torque causing the intrinsic rotation, $\Gamma^{\Omega}$ is the damping of intrinsic angular momentum and $\Gamma^v$ is the frictional damping of flow.
The fluid stress is given by
\begin{equation} 
\begin{split}
 \label{eq:ss}
	\sigma_{ij} \equiv \epsilon_{ij} \frac{\Gamma}{2} 
	\left(\Omega  - \omega\right)  - p \delta_{ij} + \eta_{ijkl} v_{kl} \\
	+ \frac{I \Omega}{2}(\partial_i v_j^*+\partial_i^* v_j)
\end{split}
\end{equation} 
to lowest nonlinear order, where $\omega \equiv \frac{1}{2} \epsilon_{ij} \partial_i v_j$ is the vorticity and $v_j^*\equiv \epsilon_{jl} v_l$ is the velocity vector rotated clockwise by $\pi/2$.
The stress is composed of the usual fluid stress terms due to the pressure $p$ and the (dissipative) viscosity tensor $\eta_{ijkl}$ present in any fluid, and two terms peculiar to chiral active fluids.
One such term is the antisymmetric stress in Eq.~(\ref{eq:ss}) proportional to $\Gamma$, which results from inter-rotor friction and couples the flow $\vv$ to the intrinsic rotations $\Omega$.
The other chiral term, $I \Omega(\partial_i v_j^*+\partial_i^* v_j)/2$, is a nonlinear contribution that arises from the conservation
of angular momentum~\cite{Banerjee2017}.

We now linearize these equations around the appropriate steady state.
We are interested in the regime in which gradients of intrinsic angular rotation $\Omega$ are negligibly small: this corresponds to a velocity $v_0$ and length scale $r_0$ such that $\Gamma/I  \ll v_0/r_0 \ll \tau/ \Gamma$ (implying that $\tau \gg \Gamma^2/ I$).
If $\Omega = \Omega_0$ is constant, the terms involving the gradients of $\Omega$ in the equation for $\vv$
vanish, and the only chiral term that remains has the form $I \Omega_0/2 \nabla^2 \vv^*$, i.e., it acts as an odd viscosity term with the value
of odd viscosity given by $\eta^o = I \Omega_0/2$.

When $\Omega$ is constant, the equations of motion for a chiral active fluid are captured by the two dynamical fields 
$\vv(\rv,t)$ and $\rho(\rv,t)$.  
For a fluid in a disk geometry, these equations can 
be solved for the steady state (defined by $\partial_t \vv_0 = \partial_t \rho_0 = 0$).
If the speed of the steady-state flow is small compared to the speed of sound ($|\vv_0|\ll c$), 
the steady-state flow can be considered incompressible, so $\nabla \cdot \vv_0 = 0$ and 
we indeed have $\partial_t \rho_0 = 0$.
The flow is then defined by the vorticity $\omega \equiv \nabla \times \vv/2$, which in the steady-state satisfies a Helmholtz equation.
For rolling (i.e., the chiral active fluid equivalent of no-slip) or partial-slip boundary conditions, the speed of chiral active particles at the edge is not zero, but 
is instead proportional to $\Omega_0$. Alternatively, the boundary conditions can be defined via a constant $\omega \sim \Omega_0$. 
Taking the limit that the surface friction $\Gamma^v$ is small, the equation for $\omega$ in the steady state becomes Laplace's equation: $\nabla^2 \omega = 0$. 
For constant boundary conditions, the solution to this equation will also be constant in space, i.e., $\omega =$ const.
This solution is precisely a rigid-body rotation, and part of the family of solutions described in Ref.~\cite{vanZuiden2016}.
This solution describes well the flow profile observed in both a continuum analytical solution and particle-based
numerical simulations of a chiral active fluid in a disk~\cite{vanZuiden2016}.
In this situation, the equations of motion in the rotating frame include a Coriolis force term, as explained in the case of the rotating fluid. 
Hence, the linearized equations describing density waves in the active fluid read
\begin{align}
\partial_t \rho(\rv,t) & = - \rho_0 \nabla \cdot \vv(\rv,t) \\
\partial_t \vv & = - c^2 \nabla \rho/\rho_0 + 2 \omega_R \vv^* + \nu^o \nabla^2 \vv^*
\end{align}
(with $\nu^o = \eta^o/\rho_0$), which are Eqs.~(1--2) of the main text.

\section{Finite-element simulations in a system with boundaries}

To test the analytical theory, we performed Finite-Element Analysis simulations
using COMSOL Multiphysics software. 
Eqs.~(1--2) of main text were simulated by modifying the
time-dependent Euler equation physics within the Acoustics module.
For these simulations, we used the modified
disk geometry shown in Fig.~3 of main text with hard-wall boundary conditions and a
small source with frequency $\omega_s$ on the left-hand side of the disk.
{These boundary conditions correspond to no penetration of the boundary, $\vv \cdot \hat{n} = 0$,
where $\hat{n}$ in combination with zero force in the transverse direction (i.e., perpendicular to $\hat{n}$). 
We have checked that qualitatively, the choice of no-slip boundary conditions (i.e., $\vv = 0$) in our simulations would not change Fig.~3, see Fig.~\ref{fig_no_slip}.
The radial profiles are fitted the analytic expressions $\ee^{- r \omega_B/c}$ for low odd viscosity, and 
$\ee^{- r c/2 \nuo} \sin \left( r \sqrt{\omega_B/\nuo} \right)$ for high odd viscosity.
}
For Fig.~3a--b, in arbitrary units, the parameters used were $(c,\omega_B,\nuo,\rho_0) = (8,40,0.1,1)$;
the disk size was defined via radius $3$ and the source was chosen with $\omega_s = 10 < \omega_B$.
For Fig.~3c--d, the parameters used were $(c,\omega_B,\nuo,\rho_0) = (15,1000,2,1)$;
the disk size was defined via radius $3$ and the source was chosen with $\omega_s = 20 \ll \omega_B$.
In figure~\ref{fig_vorticity}, we show the vorticity fields corresponding to both cases.

\begin{figure}
	\includegraphics[width=0.75\columnwidth]{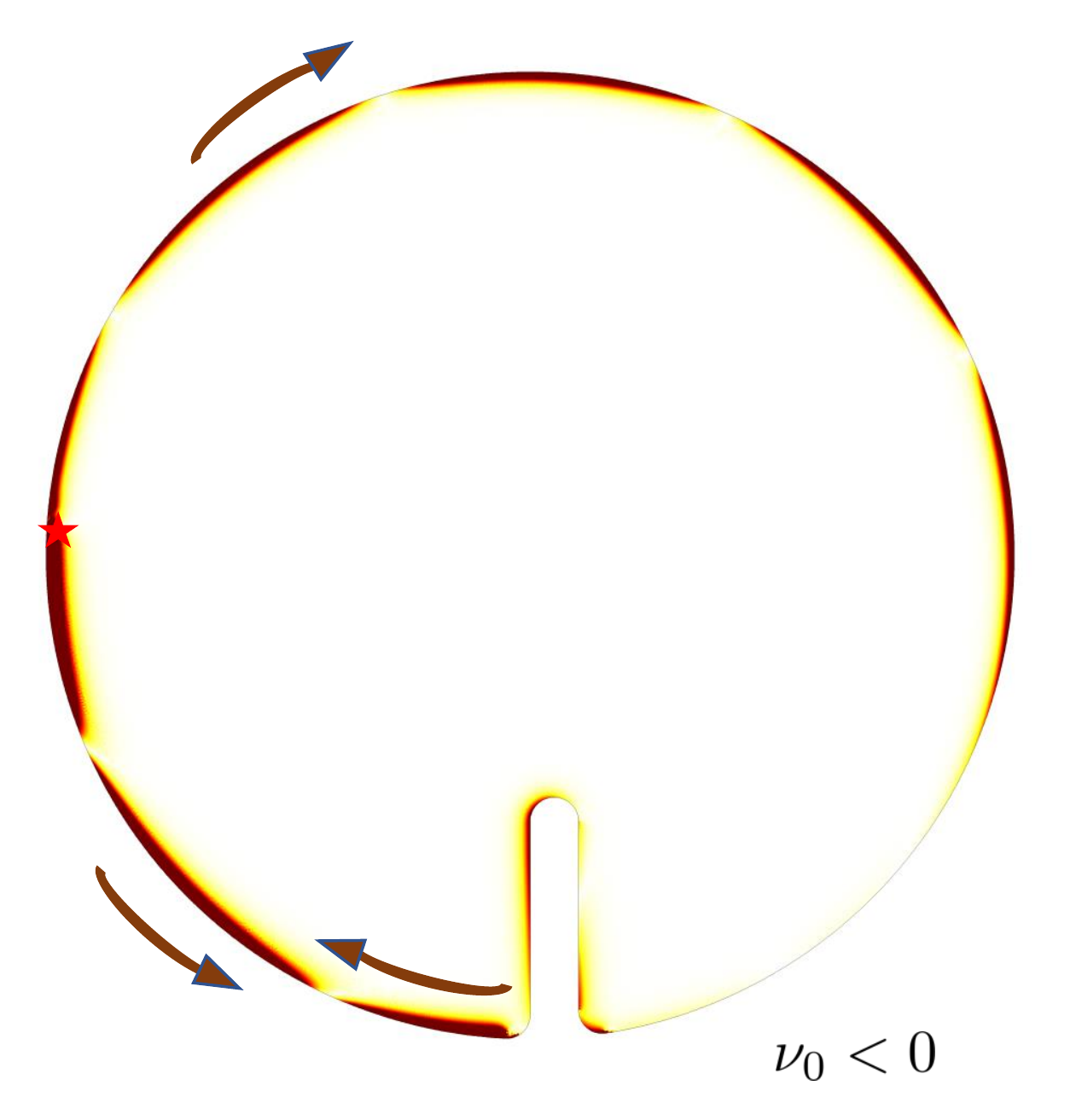}
	\caption{
{Edge states in a topologically trivial gap. 
When the sign of odd viscosity is switched from positive to negative (while keeping the value of $\omega_B$ fixed), the Chern numbers change from $\pm 2$ to $0$.  This highlights the importance of considering odd viscosity when characterizing topological density waves, for example in plasmas and chiral active fluids. In the topologically trivial case presented here, there are edge states propagating in both CW and CCW directions from the source. Furthermore, the edge waves lose topological protection and backscatter into each other, as can be seen at the notch in the bottom of the disk.
(See Supplementary Movie for time-dependent wave propagation and scattering).}}
	\label{fig_trivial_edge_states}
\end{figure}

\begin{figure*}[p]
	\includegraphics[width=1.75\columnwidth]{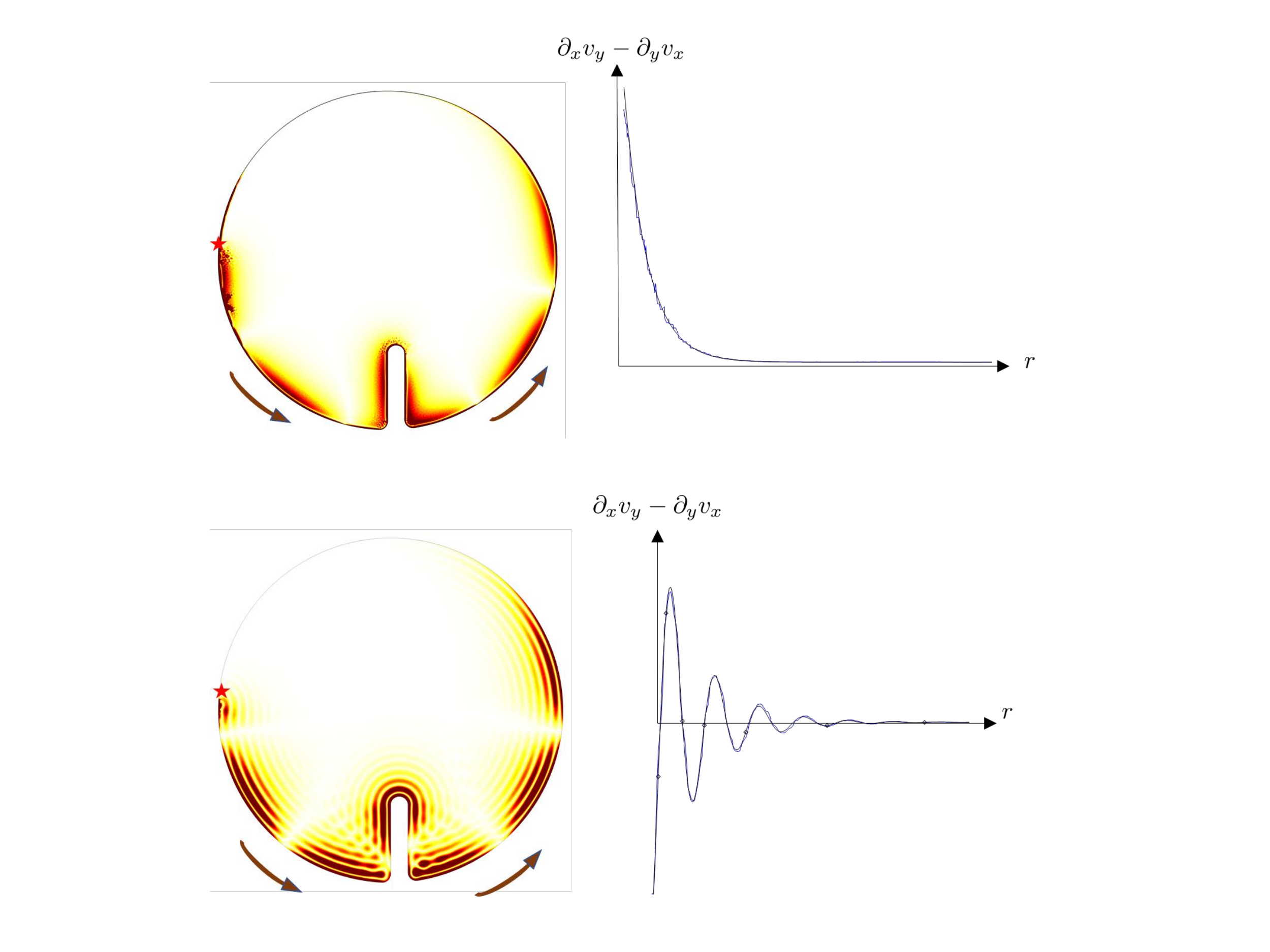}
	\caption{
{Vorticity in the topological edge state.
We see that the vorticity profile follows closely the density profile
in both low-odd-viscosity (top) and high-odd-viscosity (bottom) regimes (c.f. Figure~3 of the main text for density profiles within the same simulation).
This is expected from the eigenmode solution for the edge states,
which couples the density and vorticity fields.}
}
	\label{fig_vorticity}
\end{figure*}

\begin{figure*}[p]
	\includegraphics[width=1.75\columnwidth]{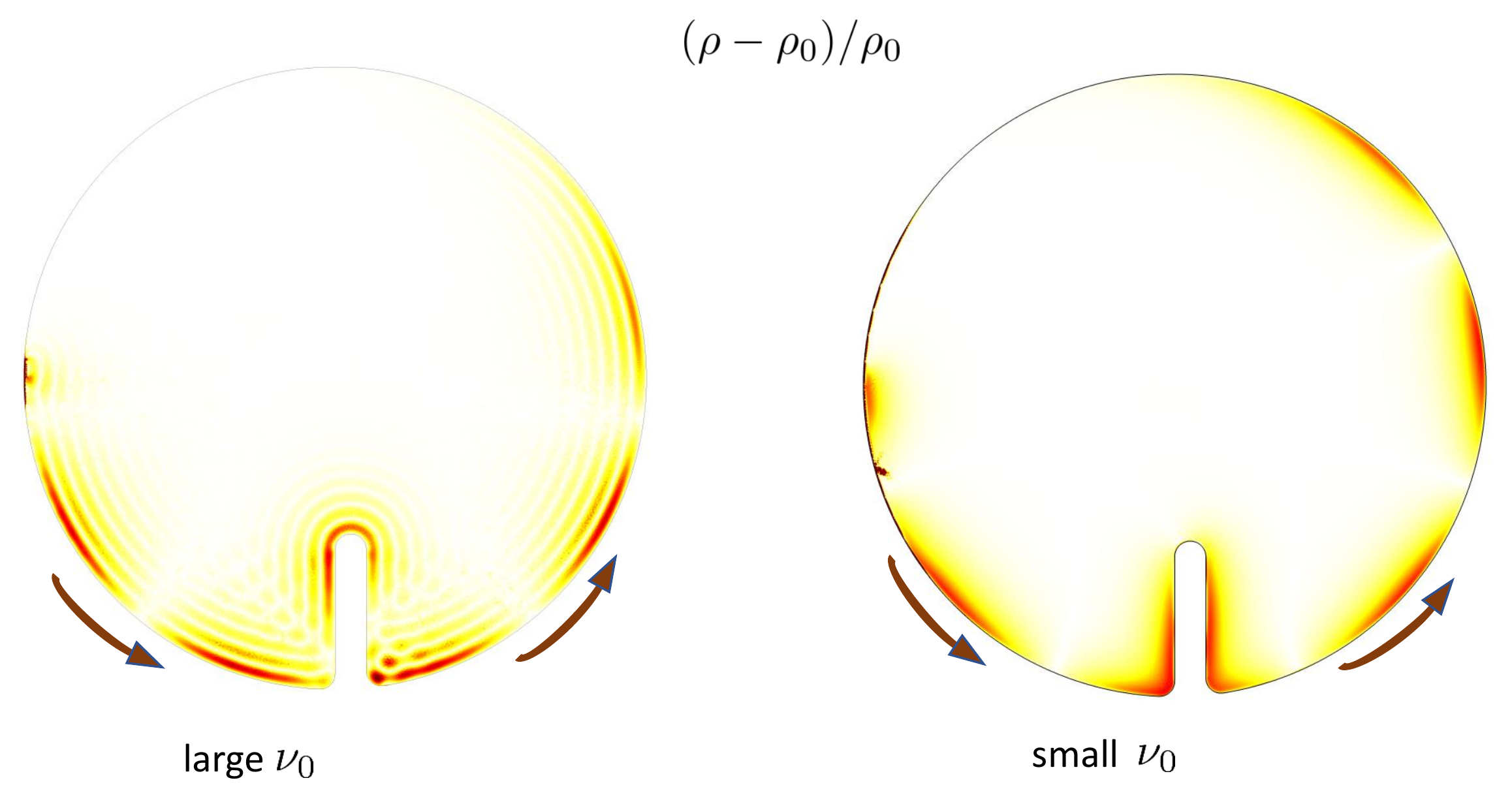}
	\caption{
{Simulations with no-slip boundary conditions. 
The edge-state snapshot and profile of the density waves do not noticeably change
when the transverse boundary condition is changed {from the boundary condition used in Figure~3 of the main text
to a no-flow boundary condition (present figure)}. This is expected from the eigenmode solution for the edge-state profile,
which does not depend on the exact boundary conditions.}
}
	\label{fig_no_slip}
\end{figure*}

When $\omega_B \nuo < 0$, the first Chern number of the bands is zero, meaning that no topologically protected edge states are expected.
We show in figure~\ref{fig_trivial_edge_states} a numerical simulation of this situation with $(c,\omega_B,\nuo,\rho_0) = (8,40,-0.1,1)$, where we observe two counter-propagating edge states, which are not topologically protected. In this case, backscattering between the edge modes clerly occurs at a defect (see also Supplementary Movie).



\section{Mode structure}
We take the Fourier transform of Eqs.~(1--2) by assuming a wavelike solution of the form 
$\ee^{\ii (\omega t - \qv \cdot \rv)}$ for $\rho$, $v_x$, and $v_y$. Dividing both sides by $\ii$ leads to
the eigenvalue problem
\begin{gather}
\omega \begin{bmatrix} \rho \\ v_x \\ v_y \end{bmatrix}
 =
  \begin{bmatrix}
   0 & \rho_0 q_x &\rho_0 q_y\\
   c^2 q_x/\rho_0 & 0 & - \ii (\omega_B - \nu^o q^2)\\
   c^2 q_y/\rho_0 & \ii (\omega_B - \nu^o q^2) & 0 \\
   \end{bmatrix}
\begin{bmatrix} \rho \\ v_x \\ v_y \end{bmatrix}
\label{eq:mtr}
\end{gather}
{The off-diagonal coupling of the form $\ii (\omega_B - \nu^o q^2)$ leads to 
a new structure of the eigenvector corresponding to the eigenvalue $\omega_0(\qv) = 0$.
This eigenvector has the form $u^0_\qv = (\ii [\omega_B - \nu^o q^2],-q_y,q_x)$. 
In other words, the chiral terms couple density and vorticity ($=\nabla\times\vv\sim-q_y v_x + q_x v_y$) within this eigenmode.}
 Note that the spectrum can be computed for the case $\nu^o = 0$,
 $\omega = \pm \sqrt{\omega_B^2 + c^2 q^2}$ and the general answer,
 Eq.~(3) found using the substitution $\omega_B \rightarrow \omega_B - \nu^o q^2$.
In this sense, the role of odd viscosity is to rescale the Lorentz force in a lengthscale-dependent manner.
Because in the limit $q \rightarrow 0$, the term $\nu^o q^2 \rightarrow 0$, 
the presence of odd viscosity in the absence of $\omega_B$ does not lead to a spectrum with a gap at $q = 0$.

\medskip
{
\section{Topological invariants}
\label{si_sec_topological_invariants}

This section is devoted to the precise definition of topological invariants in our problem. We first define the relevant quantities, and then expose an argument adapted from \cite{Volovik1988} to show that first Chern numbers for the bands are defined only when both the time-reversal breaking characteristic pulsation $\omega_B$ and the odd viscosity $\nu^o$ are nonvanishing. When $\omega_B = 0$, the gap closes, which allows a topological phase transition. More surprisingly, the gap does not close when 
$\nu^o =0 $, although in that case the Chern numbers nevertheless becomes ill-defined, and this also allows for a topological phase transition as the odd viscosity changes sign. 

A similar mechanism was analyzed in the Bogoliubov-de Gennes description of superfluid helium \cite{Volovik1988,Volovik2009b,Silaev2014}, where the inverse effective mass plays the same role as odd viscosity and allows for a regularization, and can also be considered from the point of view of Green functions \cite{Gurarie2011,Essin2011}. We also point out that such a regularization naturally occurs as a formal but necessary addition in the mathematical analysis of continuum topological insulators. This aspect was discussed in high energy physics~\cite{So1985,Coste1989,Volovik2009}, for electromagnetic continua \cite{Silveirinha2015,Silveirinha2016} and in the mathematical physics literature~\cite{Bal2018}. The marginal nature of a single Dirac (or Dirac-like) cone was also discussed by Volovik \cite[\S~11.4.2]{Volovik2009}. In the usual case of a topological insulator on a lattice, such a regularization is always present, due to the periodicity of momentum space. Interestingly, similar issues arise in condensed matter systems when, e.g., a single valley is considered~\cite{Li2010}.
}

\medskip

After a change of coordinates, the eigenvalue problem \eqref{eq:mtr} can be written
\begin{equation}
	\ii \mathcal{L}(\kv) \begin{bmatrix} \rho/\rho_0 \\ \vv/c \end{bmatrix} = \omega \begin{bmatrix} \rho/\rho_0 \\ \vv/c \end{bmatrix}
\end{equation}
where $\kv = c \qv$ and
\begin{equation}
	\mathcal{L}(\kv) = \Lv \cdot \Lambda = k_x \Lambda_x + k_y \Lambda_y + (\omega_B - (\nu^o/c^2) k^2) \Lambda_z.
\end{equation}
In this last equation, $\Lv = (k_x, k_y, \omega_B - (\nu^o/c^2) k^2)$ and $\Lambda = (\Lambda_{x}, \Lambda_{y}, \Lambda_{z})$ are the $3 \times 3$ Hermitian matrices
\begin{equation}
	\Lambda_{x} = \begin{bmatrix} 0 & 1 & 0 \\ 1 & 0 & 0 \\ 0 & 0 & 0 \end{bmatrix}
	\;
	\Lambda_{y} = \begin{bmatrix} 0 & 0 & 1 \\ 0 & 0 & 0 \\ 1 & 0 & 0 \end{bmatrix}
	\;
	\Lambda_{z} = \begin{bmatrix} 0 & 0 & 0 \\ 0 & 0 & -\ii \\ 0 & \ii & 0 \end{bmatrix}
\end{equation}
satisfying the angular momentum relation $[\Lambda_x, \Lambda_y] = \ii \Lambda_z$ (and circular permutations thereof).

In the general case in which $\Lv = (L_x, L_y, L_z)$, the spectrum of $\mathcal{L}(\kv)$ is
\begin{equation}
(\omega_{\pm}, \omega_0) = (\pm \sqrt{L_x^2 + L_y^2 + L_z^2}, 0).	
\end{equation}
The acoustic bands $\omega_{\pm}$ are gapped if and only if the vector $\Lv(\kv)$ is non-vanishing everywhere, and in this case we can consider normalizing the vector $\Lv(\kv)$ to define $\nv \equiv \Lv/\lVert\Lv\rVert$. 

A topological invariant can be defined for the bands of $\mathcal{L}(\kv)$, provided that the vector $\nv(k)$ converges to a value $\nv(|\kv| \to \infty) \to \nv(\infty)$ which is independent of the direction of $\kv$. In this case, one can assign this value $\nv(\infty)$ to a point at infinity, and consider the map $\kv \mapsto \nv(\kv)$ as defined on the compactified plane, which is equivalent to a sphere. Hence, $\nv$ can be seen as a map from the sphere $S^2$ to the sphere $S^2$. As such, it defines a compactified version of $\mathcal{L} = \Lv \cdot \Lambda$ defined as $\mathcal{N} = \nv \cdot \Lambda$, which can be seen as a map from the sphere $S^2$ to Hermitian matrices. The matrices $\mathcal{L}$ and $\mathcal{N}$ have the same eigenvectors, and therefore the same Berry curvatures, but the eigenvalues $0, \pm \omega$ of $\mathcal{L}$ are flattened to $0, \pm 1$ in $\mathcal{N}$. As $\mathcal{N}$ is defined on a compact manifold, its bands (or rather the corresponding vector bundles) have well defined first Chern numbers, which are directly related to the index (or degree) of the map $\nv : S^2 \to S^2$. By this construction, Chern numbers are attributed to the bands of the operator $\mathcal{L}(\kv)$. The first Chern numbers are topological invariants. As such, they do not change under continuous deformations of the operator $\mathcal{L}(\kv)$ (such as parameter changes), as long as (i) the gap between the bands remains open and (ii) the direction of the vector $\Lv(\kv)$ at infinity remains fixed. 

This construction can naturally be extended to more general forms of operators $\mathcal{L}(\kv)$ as follows: each band is associated to a family of orthogonal projectors $P(\kv)$ (for example, $P_{+}(\kv) = \ket{\psi_+(\kv)}\bra{\psi_+(\kv)}$) over the plane. A compactified version of this family can be defined when the projectors $P(\kv)$ have a common limit $P(|\kv| \to \infty) \to P(\infty)$ independent of the direction of $\kv$, and the first Chern number of the band is obtained by integrating the Berry curvature $\text{tr}[P \dd P \wedge \dd P]$. Again, this quantity is well-defined as long as the compactified projector family is well-defined, namely as long as (i) there is no gap closing and (ii) the projectors have a common limit at infinity.

In our case, we indeed have
\begin{equation}
	\nv(|\kv| \to \infty) \to [0,0,-\sign(\nu^o/c^2)]
\end{equation}
as long as $\nu^o \neq 0$. One immediately notices that this direction depends on the sign of odd viscosity! Hence, there must be a phase transition when $\nu^o \to 0$. In this limit (i.e., when the odd viscosity vanishes), $\nv(\kv \to \infty) \sim (k_x/k, k_y/k, 0)$ where $k^2 = k_x^2 + k_y^2$, namely $\nv$ goes to the equator of the sphere at infinity. Therefore for zero odd viscosity, it is not possible to compactify the momentum plane. As a consequence, the topological Chern number becomes ill defined for zero odd viscosity. This is what allows a topological phase transition to take place without gap closing at $\nu^o \to 0$.

Let us now define and compute the topological invariant mentioned above. We first go back to the general case. It is convenient to write $\Lv(\kv) = L (\sin \theta \cos \phi, \sin \theta \sin \phi, \cos \theta)$, so that the normalized eigenvectors are
\begin{align}
 	\psi_{\pm} &= \frac{1}{\sqrt{2}} \begin{bmatrix}
 		\sin(\theta) \\
 		\pm\cos(\phi) - \ii \cos(\theta) \sin(\phi) \\
 		\pm\sin(\phi) + \ii \cos(\theta) \cos(\phi)
 	\end{bmatrix} \\
\intertext{and}
 	\psi_{0} &=  \begin{bmatrix} 
 		\ii \cos(\theta) \\
 		-\sin(\theta) \sin(\phi) \\
 		\sin(\theta) \cos(\phi)
 	\end{bmatrix}.
 \end{align}
The Berry connections $A_{j} = \braket{\psi_{j}, \ii \, \text{d} \psi_{j}}$ are then given by $A_{\pm} = \pm \cos(\theta) \dd\phi$ and $A_{0} = 0$, and the Berry curvatures $F_{j} = \dd A_{j}$ are $F_{\pm} = \mp \sin(\theta) \dd\theta \wedge \dd\phi$ (note that there is a factor of two with respect of the more usual case of a massive Dirac cone, with Pauli matrices), and $F_{0} = 0$. In terms of cartesian coordinates, $F_{\pm} = \mp 1/2 \epsilon^{i j k} n_i \dd n_j \wedge \dd n_k$. This expression allows to express the first Chern number of the band $\pm$ as
\begin{align}
	\label{chern_number_index}
	C_{1}(\pm) &= \frac{1}{2 \pi} \int F_{\pm} \\
	&= \mp \frac{1}{2 \pi} \int_{\RR^2} \frac{\Lv}{L^3} \cdot \left( \partial_{k_x} \Lv \times  \partial_{k_y} \Lv \right) \dd k_x \dd k_y \in 2 \ZZ. \notag
\end{align}
Let us stress one more time that while the last expression involves an integral on the plane, it is pulled back from an integral on the sphere which ensures its quantization. This is only valid when $\nv(\kv)$ converges to a single value independent of the direction when $|\kv| \to \infty$. In this case, this integral is equal to \emph{two times} the index (or degree) of the map $\nv = \Lv/L : S^2 \to S^2$, so $C_{1}$ is an even integer. In the general case where this constraint is not enforced, the value of the integral is 
\emph{a priori} unconstrained, and it does not necessarily define a topological invariant.

In our case where $\Lv = (k_x, k_y, \omega_B - (\nu^o/c^2) k^2)$ we find
\begin{equation}
\begin{split}
	F_{\pm} &= \mp \frac{\omega_B + k^2 (\nu^o/c^2)}{[k^2 + (\omega_B - (\nu^o/c^2) k^2)^2]^{3/2}} \dd k_x \wedge \dd k_y \\
			&= \mp \frac{1 + \omega_B (\nu^o/c^2) \bar{q}^2}{[\bar{q}^2 + (1 - \omega_B (\nu^o/c^2) \bar{q}^2)^2]^{3/2}} 
	\dd \bar{q}_x \wedge \dd \bar{q}_y.
\end{split}
\end{equation}
Integrating over the plane yields (we assume $c^2 > 0$) we have for the lower band with dispersion $\omega_{-}$
\begin{equation}
	C_{1}(-) = \sign(\nu^o) + \sign(\omega_B).
\end{equation}
The same result can also be obtained by recognizing that equation \eqref{chern_number_index} is two times the index of the map $\nv : S^2 \to S^2$. Indeed, $\nv(0) = (0,0,\sign(\omega_B))$ while $\nv(\infty) = (0,0,-\sign(\nu^o))$, and for intermediate values $\nv$ moves towards the equator, so the map is nontrivial if and only if $\sign(\omega_B) = \sign(\nu^o)$ and exchanging the signs of the parameters changes the index to its opposite.
{
\section{Effect of ordinary viscosity}

In this section, we discuss the effects of ordinary viscosity. We will argue that (i) ordinary viscosity alone does not regularize the hydrodynamic theory, i.e. it does not allow to compactify the wave operator and to define topological invariants; (ii) when ordinary viscosity is small compared to odd viscosity, but nonzero, the main conclusions of our analysis remain unchanged, up to a probable attenuation of both bulk waves and edge waves, which is expected to be momentum-dependent; (iii) at larger values of the radio of ordinary viscosity over odd viscosity, new phenomena are expected.

\begin{figure}
	\includegraphics[width=3.4in]{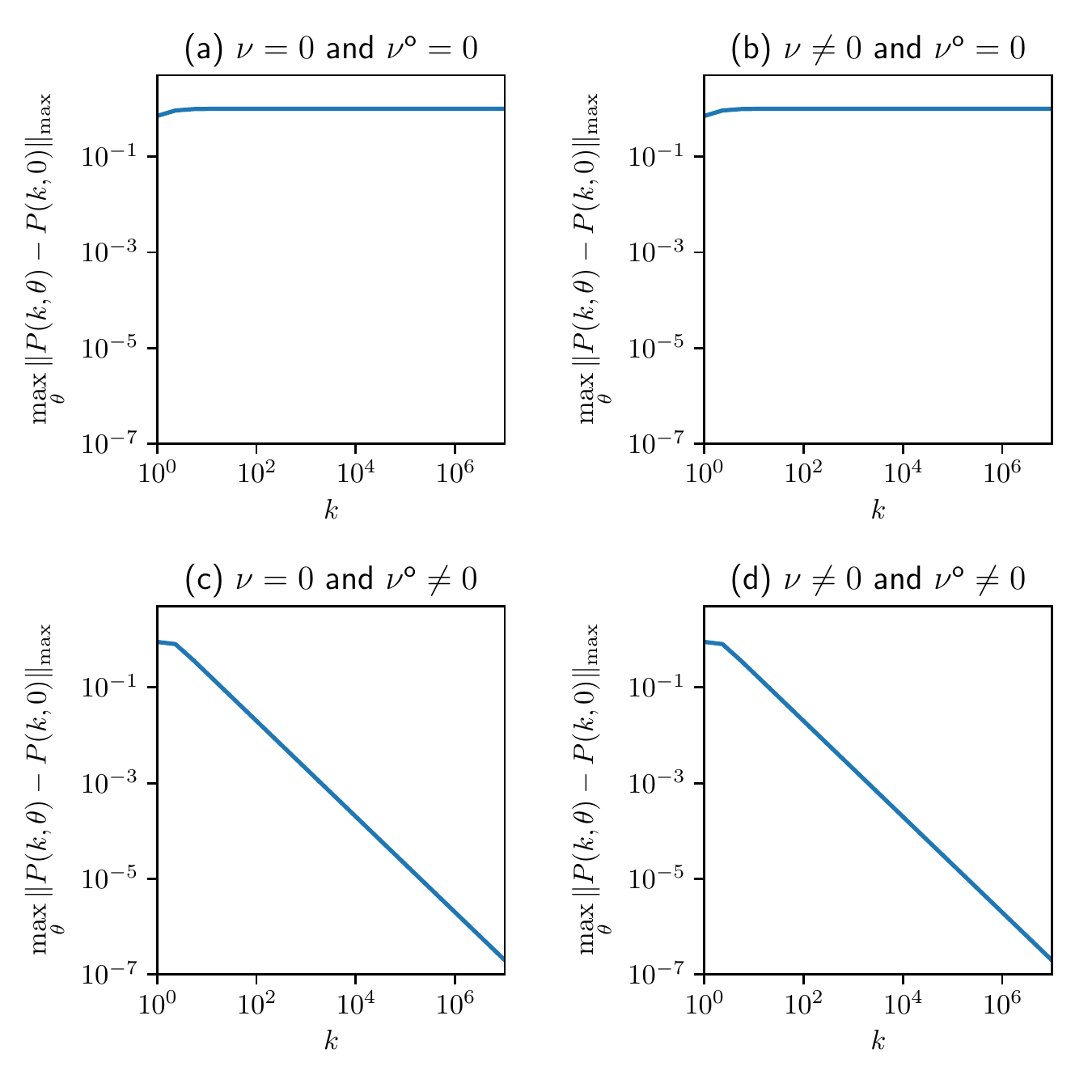}
	\caption{Angular dependency of the projector as a function of the momentum amplitude $k=|\kv|$. 
	We plot the quantity $\displaystyle\max_{\theta} \, \lVert P(k,\theta) - P(k,0) \rVert_{\max}$ as a function of $k$
	for (a) no viscosity at all (b) only ordinary viscosity (c) only odd viscosity and (d) both ordinary and odd viscosities.
	When they are nonzero, we take ordinary viscosity to be $\nu = \num{0.25}$ and odd viscosity to be $\nu^{\text{o}} = \num{0.5}$.
	}
	\label{si_projector_angular_dependency}
\end{figure}

\begin{figure}
	\includegraphics[width=3.4in]{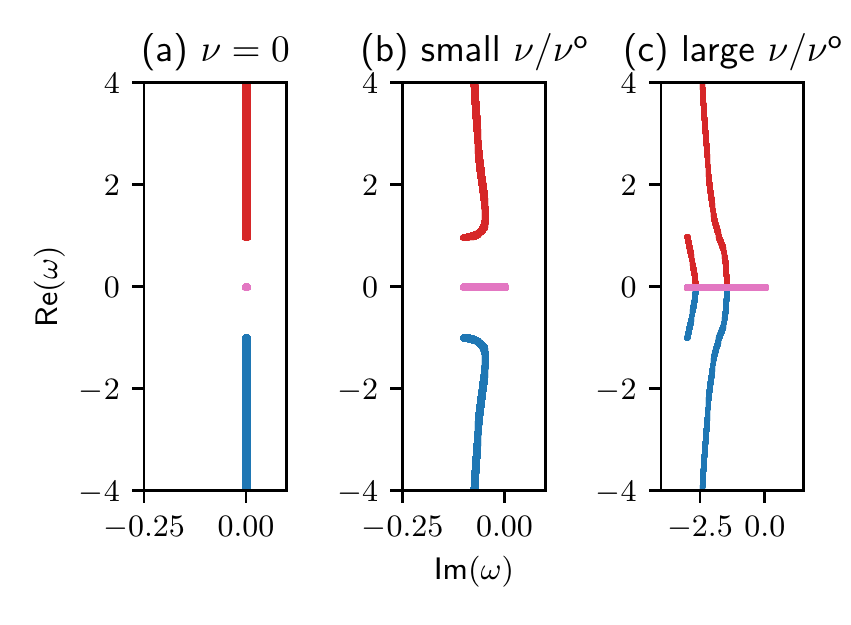}
	\caption{Spectrum of the wave operator in the complex plane for (a) no ordinary viscosity (b) small ordinary viscosity and (c) high ordinary viscosity.
	In all cases, $\omega_B = \num{1}$ and $\nu^{\text{o}} = \num{0.5}$. Ordinary viscosity is (a) $\nu = \num{0}$ (b) $\nu = \num{0.1}$ (c) $\nu = \num{3.0}$.
	The spectra of $\mathcal{L}(\kv)$ are computed for $- \num{2.5} \leq k_x,k_y \leq \num{2.5}$. 
	Higher values of $|\kv|$ correspond to points outside of the plot range, here with higher values of $|\text{Re}(\omega)|$.
	}
	\label{si_spectra_ordinary_viscosity}
\end{figure}

When ordinary viscosity is present, an additional non-Hermitian term is added to the wave operator $\mathcal{L}(\kv)$ which becomes
\begin{equation}
	\mathcal{L}(\kv) =
	\begin{bmatrix}
	0 & k_x & k_y \\ 
	k_x & - \ii (\nu/c^2) k^2 & - \ii (\omega_B - (\nu^o/c^2) k^2) \\
	k_y &  \ii (\omega_B - (\nu^o/c^2) k^2) & - \ii (\nu/c^2) k^2
	\end{bmatrix}
\end{equation}
where $\nu=\eta/\rho_0$ is the ordinary viscosity. In this section, the eigenvectors $\psi_{\pm,0}$ are labeled through the real part of their eigenvalues.

As explained in section \ref{si_sec_topological_invariants}, compactified versions of the projectors $P_{\pm,0}(\kv) = \ket{\psi_{\pm,0}(\kv)}\bra{\psi_{\pm,0}(\kv)}$ can be defined when they have a common limit at infinity. Writing $\kv = k (\cos\theta,\sin\theta)$ and $P(k,\theta) = P(\kv)$, we consider $\max_{\theta} |P(k,\theta) - P(k,0)|$ as a function of $k = |\kv|$ and numerically evaluate it. On figure \ref{si_projector_angular_dependency}, we plot this quantity (on a log-log scale) for (a) no viscosity at all i.e. $\nu = 0$ and $\nu^o = 0$ (b) only ordinary viscosity i.e. $\nu \neq 0$ and $\nu^o = 0$ (c) only odd viscosity i.e. $\nu = 0$ and $\nu^o \neq 0$ and (d) both ordinary and odd viscosities i.e. $\nu \neq 0$ and $\nu^o \neq 0$. The difference decreases with $|\kv|$ only when odd viscosity is present, showing that (a) ordinary viscosity does not regularize the theory and (b) does not hinder the regularization due to odd viscosity.

Note that when ordinary viscosity is present, but not odd viscosity, $\mathcal{L}(|\kv| \to \infty)$ converges to $\text{diag}(0, - \ii (\nu/c^2) k^2, - \ii (\nu/c^2) k^2)$ irrespective of the direction of $\kv$, but this is not enough to ensure that the compactification is possible because this property is not transmitted to the projectors $P_{\pm,0}(k)$.

We now consider the effect of a small ordinary viscosity on the bulk bands. 
We consider a system with nonzero $\omega_B$ and nonzero odd viscosity $\nu^o$. When ordinary viscosity is zero, the spectrum of the wave operator $\mathcal{L}$ is gapped and purely real [see figure \ref{si_spectra_ordinary_viscosity}(a)]. When a small ordinary viscosity is introduced, the spectrum becomes complex: it acquires a small negative imaginary part, describing the attenuation of the waves due to viscous dissipation [see figure \ref{si_spectra_ordinary_viscosity}(b) and figure \ref{si_radial_spectrum_low_ordinary_viscosity}]. Importantly, for small values of ordinary viscosity (compared to odd viscosity), the spectrum remains gapped [see figure \ref{si_spectra_ordinary_viscosity}(b)]: there are still three well-defined bands that continuously deform to the three bands $\pm,0$ when ordinary viscosity goes to zero, without gap closing. For each band, a spectral projector and a first Chern number can be assigned (there are subtleties due to the wave operator being non-Hermitian, see e.g., \cite{Shen2018}), equal by continuity to the one obtained at vanishing ordinary viscosity. Hence, one can reasonably expect that all the phenomenology discussed for zero ordinary viscosity will also occur when it is not strictly zero, up to an attenuation of all waves, probably depending on the wavenumber. At larger values ordinary viscosity (with respect to odd viscosity), different behaviors can occur, including the appearance of exceptional rings in the spectrum of the non-Hermitian wave operator (similar to \cite{Zhen2015}). In particular, the bands are no longer well-separated (see figure \ref{si_spectra_ordinary_viscosity}(c); on this figure, the merging points where the three bands touch correspond to circles in momentum space where the wave operator becomes non-diagonalizable i.e. rings of exceptional points, see also figure \ref{si_radial_spectrum_high_ordinary_viscosity}). Hence, it is not possible to directly extrapolate our results to this case, that would require a dedicated analysis.

\begin{figure}
	\includegraphics[width=2.5in]{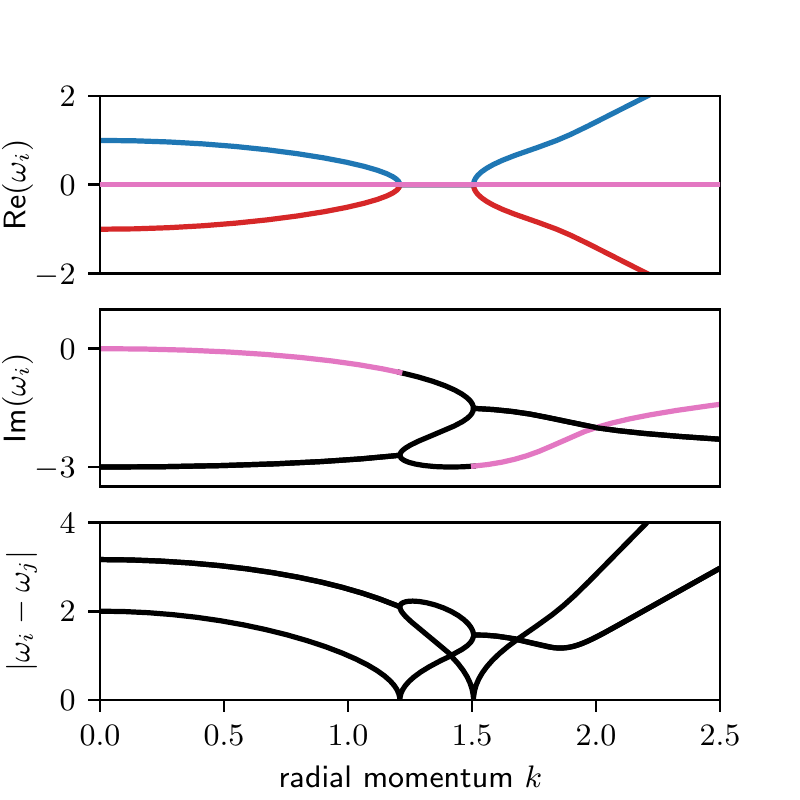}
	\caption{Spectrum of the wave operator along a radial direction for high ordinary viscosity.
	We plot the real part and imaginary part for each band, and the absolute value of the differences between the eigenvalues $|\omega_i - \omega_j|$ (that vanishes when a gap closes) with respect to the radial momentum $k$, as the eigenvalues do not depend on the angle $\theta$ of $\kv = k (\cos\theta,\sin\theta)$.
	Outside of the range of momenta where all eigenvalues have a vanishing real part, we can unambiguously define three bands $\omega_{\pm,0}(k)$ with respectively positive/negative/vanishing real part. Their real parts are respectively plotted in red/blue/pink. The imaginary part of $\omega_{\pm}$ are the same, and are plotted in black, while the imaginary part of $\omega_{0}$ is plotted in pink.
	In the range where the real parts are all vanishing, all imaginary parts are plotted in black.
	Here, $\omega_B = \num{1}$ and $\nu^{\text{o}} = \num{0.5}$ and $\nu = \num{0.1}$.
	}
	\label{si_radial_spectrum_high_ordinary_viscosity}
\end{figure}

\begin{figure}
	\includegraphics[width=2.5in]{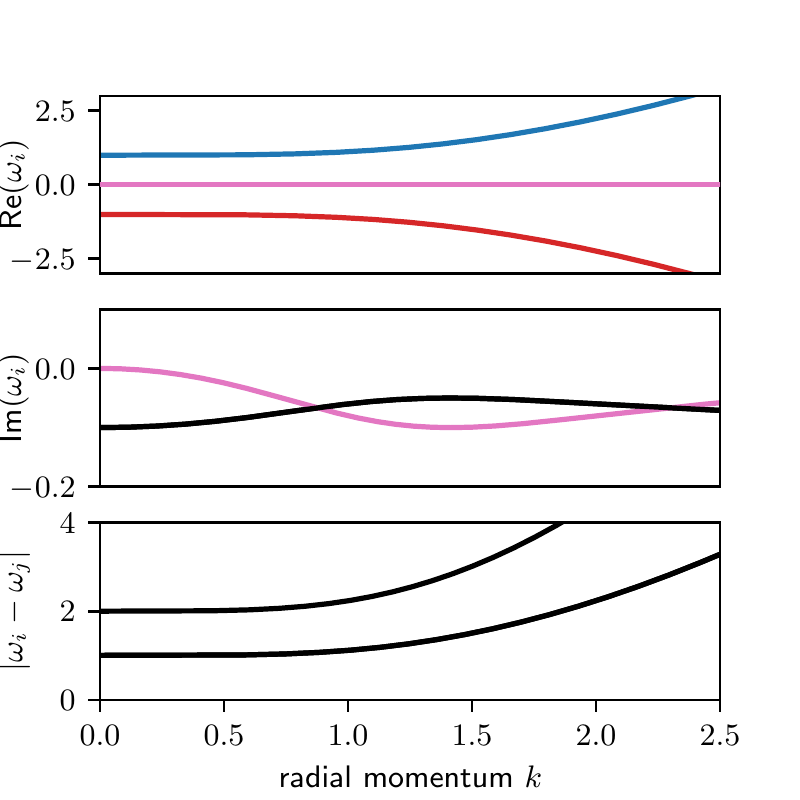}
	\caption{Spectrum of the wave operator along a radial direction for low ordinary viscosity. 
	See figure \ref{si_radial_spectrum_high_ordinary_viscosity} for details. Here, we can always distinguish the three bands $\omega_{\pm,0}(k)$.
	Here, $\omega_B = \num{1}$ and $\nu^{\text{o}} = \num{0.5}$ and $\nu = \num{3.0}$.
	}
	\label{si_radial_spectrum_low_ordinary_viscosity}
\end{figure}

}

{
\section{Shape and penetration depth of the edge states}

In the main text, the shape and penetration depth of the topological edge states are analyzed for the particular case where $(\bar{\omega},\bar{q}_x)=(0,0)$ for the edge state ($\bar{\omega}=0$ corresponds to the middle of the gap). In this section, we discuss the general case.

\begin{figure*}
	\includegraphics[width=2\columnwidth]{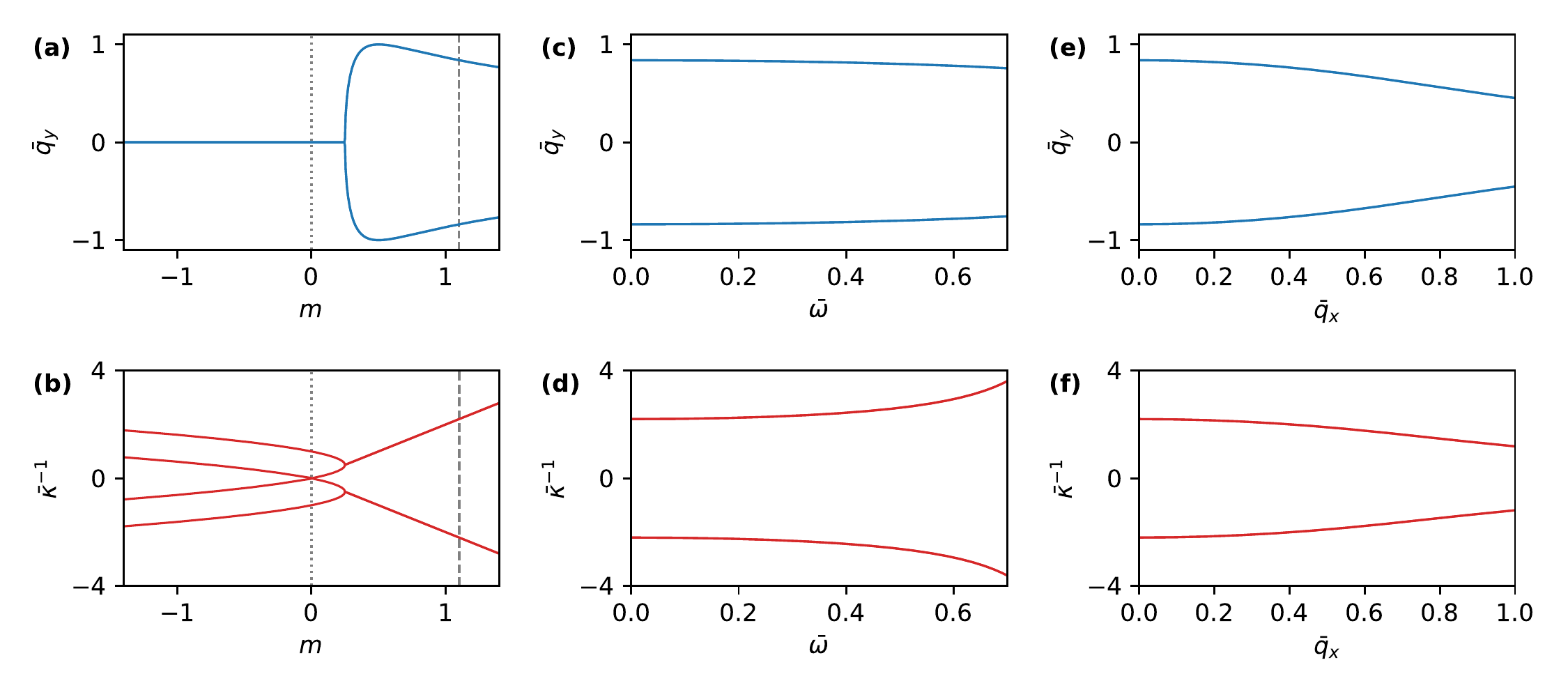}
	\caption{
{Wavevector $\bar{q}_y$ and penetration depth $\bar{\kappa}^{-1}$ of the edge states (in reduced units), (a-b) as a function of $m$ at $\bar{\omega}=0$; (c-d) as a function of the excitation (reduced) frequency $\bar{\omega}$ and (e-f) as a function of the wavevector $\bar{q}_x$ along the edge, both at fixed $m=1.1$ (this value is marked by a dashed gray line in (a-b)). In (a-b), the dotted gray line corresponds to $m=0$.}
}
	\label{si_edge_states_depth_wavenumber}
\end{figure*}

Replacing $\bar{q}^2$ by $\bar{q}_x^2 + (\bar{q}_y + \ii \bar{\kappa})^2$ in Eq.~(3) and solving for $\bar{q}_y$ as a function of all other parameters, we find four solutions $[\bar{q}_y + \ii \bar{\kappa}]_{\pm,\pm}(m,\bar{\omega},\bar{q}_x)$ that read
\begin{equation}
\pm \frac{\sqrt{2}}{2 m^2} \sqrt{- 2 m^{4} \bar{q}_x^{2} + m^{2} \pm \sqrt{4 m^{4} \bar{\omega}^{2} - 4 m^{2} + 1} - 1}
\end{equation}
where the $\pm$ are independent. In the main text, we consider the simplest case $(\bar{\omega},\bar{q}_x)=(0,0)$; the corresponding values of $\bar{q}_y$ and of the penetration depth $\bar{\kappa}^{-1}$ are plotted as a function of $m$ in figure \ref{si_edge_states_depth_wavenumber}(a,b). We also plot the same quantities as a function of $\bar{\omega}$ for fixed $m=\num{1.1}$ in figure \ref{si_edge_states_depth_wavenumber}(c,d), where it can be seen that for small values relative to the gap size at $\bar{q}=0$ (which is $2$ in units of $\bar{\omega}=\omega/\omega_B$), both the wavevector component $\bar{q}_y$ and the penetration depth $\bar{\kappa}^{-1}$ are almost constant. The same remarks hold true when $\bar{q}_x$ is real and nonzero, see figure \ref{si_edge_states_depth_wavenumber}(e,f). In the numerical simulations, the radial wavelength (corresponding to $\bar{q}_y^{-1}$) is much smaller that the azimuthal one (corresponding to $\bar{q}_x^{-1}$) with a ratio of order \num{e-2}, as seen in Figure 3 of the main text. Hence, $\bar{q}_x \ll \bar{q}_y$. In this regime, we can safely assume $(\bar{\omega},\bar{q}_x)=(0,0)$.

}
{

\section{Non-dimensionalized Navier-Stokes equation}

\def\Ma{\text{Ma}}
\def\Ro{\text{Ro}}
\def\Reodd{\text{Re}_{\text{odd}}}

The continuity equation being unmodified with respect to usual fluids, we focus on the (odd) Navier-Stokes equation. Let us define a length scale $L$ and a velocity scale $U$; a time scale is then obtained as $L/U$. Non-dimensionalizing through $\vv \to U \vv$, $\nabla \to L^{-1} \nabla$, $\partial_t \to T^{-1} \partial_t$, we obtain
\begin{equation}
	\partial_t \vv = - \Ma^{-2} \nabla \rho/\rho_0 + \Ro^{-1} \vv^* + \Reodd^{-1} \nabla^2 \vv^*
\end{equation}
where we defined the Mach, Rossby and odd Reynolds dimensionless numbers
\begin{equation}
	\Ma = \frac{U}{c}
	\qquad
	\Ro = \frac{U}{L \, \omega_B}
	\qquad
	\Reodd = \frac{U L}{\nu^o}.
\end{equation}
The odd Reynolds number is discussed in \cite{Banerjee2017}, and we call $\Ro$ the Rossby number irrespective of whether it describes a Lorentz force or a Coriolis force. Note that as shown in \cite{Banerjee2017}, a fluid with odd viscosity can be compressible even at low Mach number if the odd Reynolds number is sufficiently small, because incompressibility requires $\Ma^2/\Reodd \ll 1$.

This non-dimensionalization allows us to compare the magnitudes of the different terms. In particular, the ratio of the odd viscosity term and the Lorentz/Coriolis term is given by
\begin{equation}
	\frac{\text{odd viscosity}}{\text{Lorentz/Coriolis}} \sim \frac{\Ro}{\Reodd} = \frac{\ell_2^2}{L^2}
\end{equation}
where $\ell_2^2 = \nu^o/\omega_B$. 
Similarly, the ratio of the odd viscosity term and the compressibility term is given by
\begin{equation}
	\frac{\text{odd viscosity}}{\text{compressibility}} \sim \frac{\Ro}{\Ma^{-2}} = \Ma \frac{\ell_1}{L}
\end{equation}
where $\ell_1 = \nu^o/c$.

The lengths $\ell_2$ and $\ell_1$ separate the large length scales $L \gg \ell_i$ where odd viscosity can be neglected with respect to the other terms from the small length scales $L \lesssim \ell_i$ where it becomes important.

The dimensionless number $m$ discussed in the main text can be seen as (the square of) a reduced Mach number
\begin{equation}
	m \equiv \frac{\omega_B \nu^o}{c^2} = \frac{\Ma^2}{\Ro \, \Reodd} = \left(\frac{\ell_1}{\ell_2}\right)^2.
\end{equation}
or alternatively as the ratio of length scales $m=(\ell_1/\ell_2)^2$, which describes which term dominates odd viscosity at large length scales.

\section{Odd viscosity in various systems}

Here, we discuss the appearance of odd viscosity in several systems where parity and time-reversal symmetries are broken.

From a theoretical point of view, odd viscosity can also be understood from symmetry arguments as a hydrodynamic coefficient generically occurring in parity-violating fluids, see \cite{Jensen2012,Kaminski2014,Haehl2015} and references therein. However, this construction does not provide a value for the odd viscosity coefficient in any given system. In this section, we discuss some particular systems where the presence of odd viscosity is well-established, either experimentally or from kinetic theory.

\medskip

For gases and plasmas, theoretical results and experimental data are only available in three-dimensional systems. To connect the discussion in 2D with existing results in 3D, let us consider a quasi-2D system in the $yOz$ plane, with a magnetic field $B e_x$ orthogonal to the plane (to follow \cite{DeGrootMazur,ChapmanCowling}). In an isotropic two-dimensional system, there is only one odd viscosity. The tensor can be written $\bm{\eta}^{\text{odd}} = \eta^{\text{odd}} \, [ \bm{\sigma}_1 \otimes \bm{\sigma}_3 - \bm{\sigma}_3 \otimes \bm{\sigma}_1]$ where $\bm{\sigma}_j$ are Pauli matrices and $\otimes$ is the tensor product \cite{Avron1998}. Alternatively, the antisymmetric viscosity tensor reads in index notation
\begin{equation}
\eta^{\text{odd}}_{ijkl} = \frac{1}{2} \, \eta^{\text{odd}} \, \left(\epsilon_{ik}\delta_{jl} + \epsilon_{il}\delta_{jk}
+ \epsilon_{jk}\delta_{il} + \epsilon_{jl}\delta_{ik}\right).
\end{equation}
Hence, the only non-vanishing components are
\begin{equation}
\begin{split}
	 \eta^{\text{odd}} &= \eta^{\text{odd}}_{yyyz} = \eta^{\text{odd}}_{yyzy} = - \eta^{\text{odd}}_{yzyy} = - \eta^{\text{odd}}_{zyyy} \\
	-\eta^{\text{odd}} &= \eta^{\text{odd}}_{zzzy} = \eta^{\text{odd}}_{zzyz} = - \eta^{\text{odd}}_{zyzz} = - \eta^{\text{odd}}_{yzzz}.
\end{split}
\end{equation}

\subsection{Thermal plasmas and monoatomic gases under rotation}

The viscosity of a hot (i.e. thermal) magnetoactive  one-component plasma at strong magnetic field can be computed by kinetic theory \cite{ChapmanCowling,Landau10}. Most of the momentum in the plasma is  carried by the ions, and it is assumed that the collisions with the electrons do not contribute to the transport coefficients. 
In the same quasi-2D geometry as below, we have $\eta_{yyyz} = 2 \eta_3$ while $\eta_{yzyy} = -\eta_3$, where $\eta_3$ refers to the notations of \cite[\S~58, in particular (58.16)]{Landau10} (in this reference, the magnetic field is chosen along $z$, so the relevant components are $xxxy$ and $xyxx$). The antisymmetric part is $\eta_{yyyz}^{(A)} = 3/2 \eta_3$, so we have $\eta^{\text{odd}} \propto \eta_3$. In the hypotheses of a strong magnetic field, reference \cite[\S~59, (59.38)]{Landau10} gives
\begin{equation}
	\eta^{\text{odd}} \propto \eta_3 = \rho \frac{k_{\text{B}} T}{2 m \omega_B}
\end{equation}
where $\rho$ is the fluid density, $\omega_B = q B/m$ is the cyclotron frequency, $m$ and $q$ the masses and charges of the ions. This equation is only valid at strong magnetic field. A more general version (still for a thermal plasma) is \cite[\S~19.44 and \S~19.32]{ChapmanCowling} (see also \cite{Hooyman1954} for a sign correction)
\begin{equation}
\begin{split}
	\eta^{\text{odd}} \propto \eta_3 &= \frac{1}{2} \, \eta_{y y y z} 
	= \frac{\eta_0}{2}  \left( \frac{4 \omega_B \tau}{1 + 4 \omega_B^2 \tau^2} \right) \\
	&= \rho \frac{k_{\text{B}} T}{2 m \omega_B} \left( \frac{4 \omega_B^2 \tau^2}{1 + 4 \omega_B^2 \tau^2} \right)
\end{split}
\end{equation}
where $\tau$ is a collision time, and $\eta_0 = p \tau$ the viscosity at zero field. When $\omega_B \tau \gg 1$, the simplified case of a strong magnetic field is recovered, but here $\eta_3(B) \to 0$ when $B \to 0$. Note that the standard viscosity in the plane is also modified by the magnetic field, and \cite{ChapmanCowling}
\begin{equation}
	\eta_{yyyy} = \eta_{zzzz} 
	= \frac{-2 \eta_0}{1 + 4 \omega_B^2 \tau^2}
	= - \rho \frac{k_{\text{B}} T}{m \omega_B} \frac{\omega_B \tau}{1 + 4 \omega_B^2 \tau^2}.
\end{equation}
Hence, the ratio of odd over standard viscosity scales as $\eta^{(A)}_{y y y z}/\eta_{yyyy} = - 3/2 \omega_B \tau$.

Interestingly, the same result was found for gases (without additional internal degrees of freedom) subject to a Coriolis force \cite{Nakagawa1956}, up to the replacement $\omega_B \to 4/3 \Omega$, where $\Omega$ is the rotation rate.

\subsection{Two-dimensional electron gases, superfluids, and superconductors}

In hard condensed matter physics, odd viscosity is most often called Hall viscosity, 
and it appears in the semi-classical description of the electron fluid in a metal \cite{Alekseev2016,Pellegrino2017,Steinberg1958}
under magnetic field. There, the Hall viscosity of a two-dimensional electron gas computed from kinetic theory is
\begin{equation}
	\eta^{\text{odd}} \propto \eta_{\text{H}} = - \eta_0 \, \frac{\omega_{B} \tau_0}{1+(\omega_{B} \tau_0)^2}
\end{equation}
where $\eta_0$ is the shear viscosity of the electron fluid without magnetic field, and $\tau_0$ a collision time.
The standard viscosity under magnetic field is $\eta = \eta_0 / (1+(\omega_{B} \tau_0)^2)$ ; at large $\omega_{B} \tau_0$ the Hall viscosity dominates.
This viscosity affects the motion of electrons in the electron gas, leading in principle to measurable quantities \cite{Hoyos2012,Alekseev2016,Scaffidi2017,Pellegrino2017}. This analysis was extended to graphene \cite{Sherafati2016}, where the presence of odd viscosity has been experimentally reported in \cite{Berdyugin2018} with values as high as $\eta_{\text{H}}/\eta_0 \sim \num{0.3}$.
For example, at $T=\SI{100}{\kelvin}$, the characteristic length $\ell_2 = \sqrt{\nu^\text{o}/\omega_B} \sim \SI{1.2}{\micro\meter}$.

Hall viscosity has also been predicted in superfluids such as Helium III \cite{Volovik1984,Vollhardt,Fujii2018} and superconductors \cite{Shitade2014}, provided that time-reversal and parity are broken, a situation occurring in chiral $p + \ii p$ superfluids/superconductors.

In gapped quantum fluids, a Hall viscosity has been predicted (using adiabatic properties of the ground states under constant strains) to be topologically quantized \cite{Avron1995,Read2009,Read2011,Hoyos2012,
Hughes2011,
Hughes2013,
Bradlyn2012,
Hoyos2014,
Fremling2014,
You2016}.
In such systems, the Hall viscosity can be expressed as~\cite{Read2009,Read2011}
\begin{equation}
	\eta^{(H)} = \frac{1}{2} \, \hbar \, \overline{n} \, \overline{s}
\end{equation}
where $\overline{s}$ can be seen as (the opposite of) the average orbital spin per particle, and $\overline{n}$ is the particle number density. 

\subsection{Polyatomic gases in magnetic fields}

With the notations of \cite{DeGrootMazur} (in particular CH.~XII \S~2), used in \cite{Knaap1967}, $\eta_{yyyz} = - 2 \eta_4$ while $\eta_{yzyy} = \eta_4$. The antisymmetric part is $\eta_{yyyz}^{(A)} = - 3/2 \eta_4$, and hence $\eta^{\text{odd}} \propto - \eta_4$. For particular models of gases of non-spherical particles (e.g., polyatomic molecules) under magnetic field, kinetic theory calculations \cite{Kagan1962,Kagan1962b,Knaap1967,Kagan1967} show that 
\begin{equation}
	-\eta^{\text{odd}} \propto \eta_4 = \eta_0 \tilde{\psi} \left( 6 \frac{\tilde{\Theta}}{1+\tilde{\Theta}^2} + 4 \frac{2 \tilde{\Theta}}{1+4 \tilde{\Theta}^2} \right).
\end{equation}
In this equation, $\eta_0$ is the viscosity coefficient without magnetic field and $\tilde{\Theta}=K \mu B/\hbar p$ where $B$ is the magnetic field, $\mu$ is the magnetic moment of molecules, $p$ is the pressure of the gas, and $\tilde{\psi}$ is a dimensionless and $K$ a characteristic viscosity, both function of the microscopic parameters.
Experimental measures \cite{Korving1966,Korving1967,Hulsman1970} show that the maximum value of $\eta_4/\eta_0$ is \num{-1.81e-3} for \ce{CO} and \num{0.88e-3} for \ce{HD} (at approximately \SI{0.02}{\bar} and \SI{4}{\tesla}). (See also \cite{Beenakker1970} and references therein.)
Interestingly, the sign of $\eta_4$ (hence of odd viscosity) is controlled by the $g$-factor of the molecule through its magnetic moment, showing that the odd viscosity can be either positive or negative in this experimental realization. 
}

\section{Further discussions on plasmas}

\subsection{Three-dimensional two-component plasmas}

In two-component plasmas in the absence of external magnetic field, the heavier ions
are screened by the lighter electrons, leading to propagating ion acoustic waves.
On the other hand, a magnetic field perpendicular to a wavevector prevents screening:
the centers of electron cyclotron orbits become pinned by magnetic field lines,
suppressing motion in response to ion density waves. 
However, because ions are much heavier than electrons,
for a range of small but nonzero $\qv\cdot\Bv$,
electrons effectively screen the ions via fast motion along field lines,
whereas the ions move along cyclotron orbits.
Equations~(1--2) may be a good
description for the motion of ions in these so-called electrostatic ion cyclotron waves.
In that regime, topological edge waves would exist for 
a narrow region in three-dimensional wavevector space.
The edge waves would probably not be as robust as in the two-dimensional case, as scattering from the topological edge modes
to bulk modes in presence of a disturbance would likely possible, 
provided that the defect couples the edge modes to wavevectors outside of that narrow region.

\subsection{Unscreened plasmas}

To derive the equations that describes the Berry curvature of 
an unscreened plasma, we can consider the role of an electrostatic potential.
For a given density distribution, the electrostatic potential $\phi$
is given by the screened Poisson equation
\begin{equation}
(\nabla^2 - \lambda^{-2}) \phi = - e \rho / (\epsilon_0 M),
\end{equation}
where $\lambda$ is the screening length and $M$ is the mass of the constituent charges.
The electrostatic force on a charge $e$ is given by $- e \nabla \phi$.
Therefore, a cold plasma is described by equations
\begin{align}
\label{eq:contS}
\partial_t \rho(\rv,t) & = - \rho_0 \nabla \cdot \vv(\rv,t) \\
\partial_t \vv & = - e \nabla \phi/(M\rho_0) + \omega_B \vv^*, \\
(\nabla^2 - \lambda^{-2}) \phi &= - e \rho / \epsilon_0.
\label{eq:momS}
\end{align}
Note that in the cold plasma limit, $\nu^o \rightarrow 0$. 

Taking the Fourier transform, we can eliminate the electric potential $\phi$ via the
solution to the Poisson equation $\phi = e \rho/[\epsilon_0(q^2 + \lambda^{-2})]$.
For convenience we define the plasma frequency $\omega_p = \sqrt{\frac{\rho_0 e^2}{\epsilon_0 M^2}}$.
With this notation, the eigenvalue problem looks similar to Eq.~(\ref{eq:mtr}):
\begin{gather}
\omega \begin{bmatrix} \rho \\ v_x \\ v_y \end{bmatrix}
 =
  \begin{bmatrix}
   0 & \rho_0 q_x &\rho_0 q_y\\
   \omega_p^2 q_x/[\rho_0 (q^2 + \lambda^{-2})] & 0 & - i \omega_B\\
   \omega_p^2 q_y/[\rho_0 (q^2 + \lambda^{-2})] & i \omega_B & 0 \\
   \end{bmatrix}
\begin{bmatrix} \rho \\ v_x \\ v_y \end{bmatrix}
\end{gather}
Note that for small screening lengths, $\lambda \rightarrow 0$, 
this reduces to the same form as Eq.~(\ref{eq:mtr}), with the speed $\lambda \omega_p$
taking the place of the speed of sound $c$.
On the other hand, for unscreened plasmas, $\lambda \rightarrow \infty$,
and the effective speed of sound goes as $\omega_p^2/q^2$.
The spectrum then has a simple form $\omega^2 = \omega_B^2 + \omega_p^2$.
In the unscreened limit we use the eigenvector solutions to find the expression 
for Berry curvature:
\begin{equation}
F_\qv^{\pm} = \pm L^2 \frac{4 \frac{\omega_B}{\omega_p} 
\sqrt{1 + \left(\frac{\omega_B}{\omega_p}\right)^2}}{(1 + 2 \omega_B^2/\omega_p^2 + q^2 L^2)^2}.
\end{equation}
Note that because an unscreened plasma has no intrinsic lengthscale, an
arbitrary length $L$ must be introduced. 
{Integrating this expression over $q$-space, we find \begin{equation}
\int F_\qv^{\pm} \frac{d^2\qv}{2\pi} = \pm \frac{2 \frac{\omega_B}{\omega_p} 
\sqrt{1 + \left(\frac{\omega_B}{\omega_p}\right)^2}}{1 + 2 \omega_B^2/\omega_p^2},
\label{eq:b}
\end{equation}
which is independent of the choice of 
$L$ and corresponds to non-integer integrated Berry curvature.
This expression has two natural limits: (i) $\omega_p \ll \omega_B$ for which Eq.~(\ref{eq:b}) reduces to $\pm 1 \mp \omega_p^4/(8 \omega_B^4)$
when the gap is dominated by the topological band theory (with ${\cal C}_\pm = \pm 1$ [for $\nu^o = 0$]) 
with small
corrections due to the plasma frequency;
(ii) $\omega_p \gg \omega_B$ for which Eq.~(\ref{eq:b}) reduces to $\pm 2 \omega_B/\omega_p$.
For a non-extensive system with unscreened long-range interactions
one would expect that a band-structure analysis may be subtle.
In this light, the lack of a well-defined Chern number is a significant
feature highlighting a unique aspect of unscreened plasma physics.}

\bibliography{topological_waves_odd_viscosity}
\bibliographystyle{apsrev4-1}

\end{document}